\documentclass{emulateapj}
\usepackage[usenames]{color}
\usepackage{mathrsfs}
\usepackage{graphicx}
\usepackage{natbib}
\usepackage{enumitem}
\usepackage{soul}

\newcommand{\vvv}{{\sigma}}

\newcommand{\jvec}{{\vec{j}}}
\newcommand{\vvec}{{\vec{v}}}

\newcommand{\Bvec}{{\vec{B}}}
\newcommand{\Evec}{{\vec{E}}}

\newcommand{\gvec}{{\vec{g}}}
\newcommand{\er}  {{\vec{e}_r^{}}}

\newcommand{\etheta} {{\vec{e}_\theta^{}}}

\newcommand{\ephi}{{\vec{e}_\phi^{}}}

\newcommand{\muzero}{{}}

\newcommand{\kms}{{\mathrm{\, km \,s^{-1}}}}

\newcommand{\rperp}{{\varpi}}
\newcommand{\Bperp}{{B_{\varpi}^{}}}
\newcommand{\sa}{\hbox{sign}(A)}

\DeclareGraphicsExtensions{.pdf,.png,.jpg}
\newcommand{\alphaff}{{\alpha_{ff}}}

\usepackage{newtxtext,newtxmath}

\usepackage[colorlinks = true,
            urlcolor  = blue,
            citecolor = blue,
            anchorcolor = blue]{hyperref}

\shorttitle{Self-Similar Approach for Rotating MHD Structures}
\shortauthors{Luna, Priest \& Moreno-Insertis}

\begin{document}

\title{Self-Similar Approach for Rotating Magnetohydrodynamic Solar and Astrophysical Structures}

\author{M. Luna\altaffilmark{1,2}, E. Priest\altaffilmark{3}, \& F. Moreno-Insertis\altaffilmark{1,2}}

\altaffiltext{1}{Instituto de Astrof\'{\i}sica de Canarias, E-38200 La Laguna, Tenerife, Spain}
\altaffiltext{2}{Departamento de Astrof\'{\i}sica, Universidad de La Laguna, E-38206 La Laguna, Tenerife, Spain}
\altaffiltext{3}{Mathematics Institute, University of St Andrews St Andrews KY16 9SS, UK}




\begin{abstract}
Rotating magnetic structures are common in astrophysics, from vortex tubes and tornados in the Sun all the way to jets in different astrophysical systems. The physics of these objects often combine inertial, magnetic, gas pressure and gravitational terms. Also, they often show approximate symmetries that help simplify the otherwise rather intractable equations governing their morphology and evolution. Here we propose a general formulation of the equations assuming axisymmetry and a self-similar form for all variables: in spherical coordinates $(r,\theta,\phi)$, the magnetic field and plasma velocity are taken to be of the form: ${\bf B}={\bf f}(\theta)/r^n$ and
  ${\bf v}={\bf g}(\theta)/r^m$, with corresponding expressions for the scalar variables like pressure and density. Solutions are obtained for potential, force-free, and non-force-free magnetic configurations. Potential-field solutions can be found for all values of~$n$. Non-potential force-free solutions possess an azimuthal component $B_\phi$ and exist only for $n\ge2$; the resulting structures are twisted and have closed field lines but are not collimated around the system axis. In the non-force free case, including gas pressure, the magnetic field lines acquire an additional curvature to compensate for an outward pointing pressure gradient force. We have also considered a pure rotation situation with no gravity, in the zero-$\beta$ limit: the solution has cylindrical geometry and twisted magnetic field lines. The latter solutions can be helpful in producing a collimated magnetic field structure; but they exist only when $n<0$ and $m<0$: for applications they must be matched to an external system at a finite distance from the origin.
\end{abstract}

\keywords{magnetic fields --- magnetohydrodynamics (MHD) --- plasmas --- Sun: magnetic fields --- Sun: atmosphere}

\section{Introduction}\label{sec:intro}
Rotating magnetic structures are very common in astrophysics and solar physics. 
Rotating astrophysical jets of many kinds may be produced
\citep[e.g.,][]{ferrari2011,smith2012} following accretion processes in compact
objects such as white dwarfs, neutron stars and black holes. Stars that can
produce rotating jets include pulsars, cataclysmic variable stars, X-ray
binaries and gamma-ray bursters \citep{gracia2009}. Stellar-mass black holes
can produce microquasars \citep{blundell2011}. Other jets in star-forming
regions are associated with T Tauri stars and Herbig-Haro objects, where the
jets interact with the interstellar medium. Bipolar jets are also found with
protostars \citep{tsinganos2009} or evolved post-AGB stars and planetary
nebulae. Indeed, weak jets occur in many binary systems.  The largest and
most active jets \citep{boettcher2012,fabian2014} are created by super
massive black holes in the centres of active galaxies such as quasars and
radio galaxies and within galaxy clusters. Such jets can extend millions of
parsecs in length.

In the Sun photospheric vortex tubes are a natural occurrence in convection
simulations at downdraft junctions of cells
\citep{cattaneo2003,nordlund2009}.  They have been discovered in Swedish
Solar Telescope and SUNRISE movies of the photosphere near magnetic elements
\citep{bonet2008,bonet2010,steiner2010}, with lifetimes of 5--10 mins and
rotation periods of about 30 min.  They occur preferentially at the edges of
mesogranules and supergranules and are associated with strong magnetic fields
\citep{requerey2017}.

In the solar atmosphere, vortex tubes exist with a large range of sizes and
lifetimes, just as for magnetic flux tubes. A twisting erupting prominence is an example of a large vortex tube with a length of
60--600 Mm. Rotating macrospicules (or ``macrospicule tornadoes") are jets of
chromospheric plasma often seen in coronal holes and have an intermediate
size with a length of 4--40 Mm. Surges are large, cool ejections, often occurring at the leading edge of an active region and
reaching heights of 20--100 Mm. Rotational motions are often seen in them \citep[see e.g.,][]{gu1994,canfield1996,jibben2004}.
Type II spicules are much smaller jets of chromospheric plasma ejected from
supergranule boundaries with lengths of 2--10 Mm and torsional motions of
15--20 $\kms$ \citep{depontieu2012,depontieu2014}. They are examples of
tiny vortex tubes and are sometimes called ``spicule tornadoes".  
Vertical rotating structures have also been observed in prominence feet
called ``barb tornadoes''
\citep[e.g.,][]{orozco-suarez2012,su2012,su2014,levens2015}. The observations
show that the cool barb and surrounding hot coronal plasma are rotating with
velocities of 5--10 $\kms$ persistently for several hours.
The relationship between photospheric vortex tubes and the different kinds of tornado is not yet clear.  
The continuation of such a tube up into the chromosphere and corona to give a macrospicule or barb tornado has, however, been detected with the Swedish Solar Telescope and the Solar Dynamics Observatory \citep{Wedemeyer2009,wedemeyer2012,srivastava2017}, but it is not known whether that is a general feature and whether it applies to other kinds of tornado.

In view of the ubiquity of magnetized rotating structures in the Sun and
wider universe, it is of interest to develop analytical solutions to the MHD equations that
can describe different aspects of such
structures. Analytical studies provide insights into the basic structures and solutions that complement what can be found from numerical simulations. On the one hand, the analytical approach includes some of the key physical processes and is able to deduce the way the results depend on the dimensionless parameters. It is also able to act as a guide for and check on numerical experiments. The latter, however, have the advantage of being able to include more physical effects, but the disadvantage of being able to be run for a limited range of parameters. \cite{tsinganos1992}, \cite{tsinganos2006} and \cite{tsinganos2007,tsinganos2010}
have developed a whole range of solutions for rotating
stars. \cite{lynden-bell1994} discovered self-similar solutions that are
force-free, steady state ($\partial/\partial t=0$), and axisymmetric
($\partial/\partial \phi=0$) in terms of spherical polar coordinates
($r,\theta,\phi$), such that the vertical axis $\theta=0$ is the axis of
symmetry. They also assumed that all field lines form loops that start and
end at the origin, which could be a compact object or a concentrated source
of magnetic flux. 

The goal of the present paper is to provide a general formulation of the
equations that describe a self-similar rotating MHD system including the
effects of flow, pressure gradient and magnetic forces and gravity and to
find first solutions for them. In this work, the self-similar approach is assumed in order to simplify the full set of MHD equations, so that analytical solutions are possible. The equations are obtained by adopting the
general shape $f(\theta)/r^n$ for the 
plasma velocity, the magnetic field components, the pressure and the density, 
 with both the power ($n$) and the angular dependence
being different for each physical variable.  Our formulation, therefore,
considerably generalizes the approach of \citet{lynden-bell1994}.  Various
solutions are found ranging from potential cases, through force-free field
solutions, and to non-force-free ones for which, in addition to the magnetic
forces, either the inertial forces associated with rotation or the gas
pressure gradient are included. In the paper, we first derive the general
equations adequate for an axisymmetric, rotating MHD system with gravity
(Section~\ref{sec:basic_equations}), thereafter imposing the self-similar
ansatz (Section~\ref{sec:self-similar}). This is
followed by discussion of the potential 
(Section~\ref{sec:potential-field}) and force-free solutions
(Section~\ref{sec:fff}). The final section is devoted to non-force-free
cases (Section~\ref{sec:pure-rotation}), in particular to cases that
include gas pressure gradients (Section~\ref{sec:fff+gaspressure}) and pure
rotation \ref{sec:lowtemperature}), thus disregarding poloidal flows and the
gravity.

\section{Basic equations}\label{sec:basic_equations}
In this work, we consider stationary axisymmetric magnetic structures with
the axis of symmetry in the vertical $z$-direction. The magnetic field and
plasma velocity possess components in vertical planes through the $z$-axis
and also in the azimuthal $\phi$-direction about it. The structure is
governed by the steady-state ideal-MHD equations with $\partial / \partial t
= 0$, namely,
\begin{eqnarray} \label{eq:divb}
0 &=& \nabla \cdot \Bvec \, , \\ \label{eq:divrhov}
0 &=& \nabla \cdot \left( \rho\, \vvec\right) \, , \\ \label{eq:induction}
0 &=&  \nabla \times (\vvec\times \Bvec) \label{eq:momentum}
~,\label{eq:mhd1} \\
0 &=&  - \rho (\vvec \cdot \nabla)
\vvec -\nabla p + \frac{1}{\mu_0}(\nabla \times \Bvec)\times
\Bvec+\rho\, \vec{g}\, , \label{eq:mhd2} 
\end{eqnarray}
where $\rho$, $\vvec$, $\Bvec$, $p$ and $\vec{g}= - g\, \vec{e}_z$ are the
plasma density, the velocity, magnetic field, plasma pressure and gravity,
respectively {\bf being $g$ constant}. In the following, we use spherical coordinates
$\left(r,\theta,\phi \right)$ with $\theta=0$ along the vertical
axis. 

In agreement with the self-similar nature that we seek in this paper, we now
write dimensionless versions of the foregoing equations by introducing units
of length, $L$, time, $\tau$, and magnetic field strength, $B_0$. The units
of pressure and density can therefore be chosen as $B_0^2/\mu_0$ and
$L_0^{-2}\,\tau^2\, B_0^2/\mu_0 $. Finally, the magnetic field in the
equations is written in terms of $\vec{B}/\sqrt{\mu_0}$. We then rewrite
Equations~(\ref{eq:divb}) through (\ref{eq:mhd2}) in terms of the
dimensionless variables (e.g., by adding a hat to them) and then drop the
hats, since there should be no confusion. After carrying out these operations
Equations (\ref{eq:divb}) -- 
(\ref{eq:mhd1}) retain exactly the same shape and Equation~(\ref{eq:mhd2})
becomes: 

\begin{equation}
0 =  - \rho\, (\vvec\, \cdot \nabla)\,\vvec \;-\; 
\nabla p \;+\; (\nabla \times \Bvec)\times \Bvec \;+ \;
\rho\, \vec{g}\, ,  \label{eq:mhd2b} 
\end{equation}

\noindent with $\gvec$ now the dimensionless gravity. When $\gvec$ is
uniform, its magnitude is the only external parameter appearing in the
equations. In the rest of 
the paper we will use the dimensionless variables and equations
exclusively. The dimensionless electric field $\Evec$ and 
current $\jvec$, in particular, are defined such that they fulfill the
dimensionless ideal Ohm's and Amp\`ere's law, namely:

\begin{eqnarray}\label{eq:ohm}
\Evec = - \vvec\, \times\, \Bvec\;,\\
\noalign{\vspace{2mm}}\label{eq:ampere}
\jvec = \nabla\,\times\,\Bvec \;. 
\end{eqnarray}
 
Both the plasma velocity and magnetic field can be naturally decomposed into 
\emph{poloidal} and \emph{toroidal} components, 
\begin{eqnarray} \label{eq:vdecomp-polar-toroidal}
\vvec &=& \vvec_p + \vvec_t=\vvec_p + v_\phi\,\ephi ~,\\ \label{eq:bdecomp-polar-toroidal}
\Bvec &=& \Bvec_p + \Bvec_t =\Bvec_p +B_\phi\,\ephi ~,
\end{eqnarray}
with $\vvec_p\cdot\ephi=\Bvec_p\cdot\ephi = 0$, $\vvec_t=v_\phi\,\ephi$, and $\Bvec_t=B_\phi\,\ephi $.
The poloidal components are thus contained in $\phi=\mathrm{constant}$
planes, whereas the toroidal parts are just the azimuthal component of the
vectors. 
For axisymmetric situations $\partial/\partial \phi=0$, a natural way of defining the angular velocity
$\Omega(r,\theta)$ is by:
\begin{equation}\label{eq:azimuthal-vel-field}
 \vec{v}_t = r\, \sin \theta \, \Omega(r,\theta)\,\ephi
\end{equation}

Let us now write the equations in terms of their components and derivatives in spherical coordinates. Equations (\ref{eq:divb}) and (\ref{eq:divrhov}) become
\begin{equation}\label{eq:divb-components}
\frac{1}{r^2} \frac{\partial (r^2 \, B_r)}{\partial r} +\frac{1}{r \sin \theta} \frac{\partial (\sin \theta B_\theta)}{\partial \theta} =0 \, ,
\end{equation}
and 
\begin{equation}\label{eq:divrhov-components}
\frac{1}{r^2} \frac{\partial (r^2  \, \rho \, v_r)}{\partial r} +\frac{1}{r \sin \theta} \frac{\partial (\sin \theta \rho \, v_\theta)}{\partial \theta} =0 \, .
\end{equation}
The electric field can also be naturally decomposed into poloidal and toroidal components,
\begin{equation}
\vec{E} =  - \vvec\times \Bvec = \vec{E}_p + E_\phi\,\ephi \, ,
\end{equation}
with
\begin{eqnarray}\label{eq:electricfield-toroidal}
E_\phi\,\ephi &=& - \vvec_p \times \Bvec_p \\ \label{eq:electricfield-poloidal}
\vec{E}_p &=&-\vvec_p \times \Bvec_t - \vvec_t \times \Bvec_p \, .
\end{eqnarray}
As \citet{mestel1961} stresses, in an axisymmetric and stationary situation there is no toroidal electric field, since otherwise the $r$ and $\theta$ components of the induction Equation (\ref{eq:induction}) would imply
\begin{equation}
E_\phi = \frac{C}{r \sin \theta} \, ,
\end{equation}
which is singular at the origin except when $C=0$.  This condition implies,
from Equation (\ref{eq:electricfield-toroidal}), that the poloidal components
of velocity and magnetic field must be parallel (or zero), and that the
following relation:
 \begin{equation}\label{eq:induction-poloidal}
 E_\phi = v_r \, B_\theta -B_r \, v_\theta =0 \, 
 \end{equation}
must be fulfilled. 
The induction Equation (\ref{eq:induction}) can then be written as
\begin{equation}\label{eq:induction_2}
-\nabla \times \vec{E}_p = \nabla \times \left( \vvec_p \times \Bvec_t + \vvec_t \times \Bvec_p \right) =0 \, ,
\end{equation} or, using Equation (\ref{eq:azimuthal-vel-field}), 
 \begin{equation}
 \frac{1}{r} \vvec_p \cdot \nabla (r^2 \Omega) + \frac{1}{\muzero r} \Bvec_p \cdot \nabla (r B_\phi)=0 \, ,
 \end{equation}
which gives a relation between $\Omega$ and $B_\phi$ following a field line.
With this relation we can find a conservation law along the field involving
$\Omega$ and $B_\phi$, which is well known for
axisymmetric rotating structures \citep[see e.g.,][]{lovelace1986}. Despite
the relevance of conservation laws, we are here more interested in explicit relations between the field components, so we rewrite
Eq.~(\ref{eq:induction_2}) as:
 \begin{equation}\label{eq:induction-toroidal}
\frac{\partial }{\partial r} (r B_r v_\phi - r B_\phi v_r) + \frac{\partial }{\partial \theta} (B_\theta v_\phi -B_\phi v_\theta)=0 \, .
 \end{equation}

The momentum Equation (\ref{eq:momentum}) may be written in terms of the
vector components as:
\begin{widetext}
\begin{eqnarray} \label{eq:mom-r}
-\rho \left(v_r \frac{\partial v_r}{\partial r}  + \frac{v_\theta}{r} \frac{\partial v_r}{\partial \theta} - \frac{v_\theta^2+v_\phi^2}{r}	\right) -\frac{B_\theta}{\muzero r} \frac{\partial (r B_\theta)}{\partial r} + \frac{B_\theta}{\muzero r}\frac{\partial B_r }{\partial \theta} - \frac{B_\phi}{\muzero r} \frac{\partial (r B_\phi)}{\partial r} -\frac{\partial p}{\partial r}-\rho \, g \, \cos \theta &=&0, \\ \label{eq:mom-theta}
- \rho \left( v_r \frac{\partial v_\theta}{\partial r}	+\frac{v_\theta}{r}\frac{\partial v_\theta}{\partial \theta} +\frac{v_\theta v_r}{r} - \frac{\cos \theta }{r \sin \theta} v_\phi^2	\right) - \frac{B_\phi}{\muzero r \sin \theta}\frac{\partial \sin \theta B_\phi}{\partial \theta} + \frac{B_r}{\muzero r} \left( \frac{\partial r B_\theta}{\partial r} - \frac{\partial B_r}{\partial \theta} \right) -\frac{1}{r}\frac{\partial p}{\partial \theta} +\rho \, g \, \sin \theta &=& 0, \\ \label{eq:mom-phi}
-\rho \left( v_r \frac{\partial v_\phi}{\partial r}	+	\frac{v_r v_\theta}{r}	+\frac{v_\theta}{r \sin \theta}\frac{\partial \sin \theta v_\phi}{\partial \theta}\right) + \frac{B_\theta}{\muzero r \sin \theta }\frac{\partial \sin \theta B_\phi}{\partial \theta} + \frac{B_r}{\muzero r} \frac{\partial r B_\phi}{\partial r} ~~~~~~ ~~~~~~~~&=& 0.
\end{eqnarray}
\end{widetext}
From Equations (\ref{eq:divb}) and (\ref{eq:divrhov}) the poloidal components
of $\Bvec$ and $\rho \, \vvec$ can be written in terms of potential functions
$\hat{A}(r, \theta)$ and $\hat{\Psi}(r, \theta)$, respectively,  as
\begin{eqnarray} \label{eq:magnetic_potential}
\Bvec_p &=& \nabla \times \left( \frac{\hat{A}}{r \sin \theta} \, \ephi \right) \, ,\\
\label{eq:momentum_potential}
\rho \, \vvec_p &=& \nabla \times \left( \frac{\hat{\Psi}}{r \sin \theta} \, \ephi \right) \, ,
\end{eqnarray}
whose spherical coordinate components are
\begin{eqnarray}\label{eq:potential-A-r}
B_r &=& \frac{1}{r^2 \sin \theta} \frac{\partial \hat{A} }{\partial \theta}\, ,\\ \label{eq:potential-A-theta}
B_\theta &=& \frac{-1}{r \sin \theta} \frac{\partial \hat{A} }{\partial r} \, , \\ \label{eq:potential-psi-r}
\rho v_r &=& \frac{1}{r^2 \sin \theta} \frac{\partial \hat{\Psi} }{\partial \theta} \, ,\\ \label{eq:potential-psi-theta}
\rho v_\theta &=& \frac{-1}{r \sin \theta} \frac{\partial \hat{\Psi} }{\partial r} \, .
\end{eqnarray}

From Equation~(\ref{eq:magnetic_potential}), or from
  Equations~(\ref{eq:potential-A-r})-(\ref{eq:potential-A-theta}), 
 we see that the projected field
  lines on the poloidal planes are just the lines $\hat{A}=$ const.

\section{Self-similar solutions}\label{sec:self-similar}
We seek self-similar solutions for the physical variables with the following
dependence on distance ($r$) from the origin and angle ($\theta$):
\begin{equation} \label{eq:selfsimilardefinitions}
\Bvec = \frac{\widetilde{\Bvec}(\theta)}{r^n} \,, \,\, \vvec = \frac{\widetilde{\vvec}(\theta)}{r^m}\, ,\,\, \rho = \frac{\widetilde{\rho}(\theta)}{r^q} \, ,\,\, p=\frac{\widetilde{p}(\theta)}{r^s} \, ,
\end{equation}
where, at this stage, the constants $n$, $m$, $q$, $s$ could be positive or negative. It is important to note that we are assuming self-similarity in order to reduce the complexity of the MHD Equations (\ref{eq:mom-r})-(\ref{eq:mom-phi}). Many astrophysical and solar systems (see Sec. \ref{sec:intro}) have large-scale complex structures different from the self-similar shape. However, localized regions of those structures can be described by the self-similar fields of Equation (\ref{eq:selfsimilardefinitions}). In addition, solutions with $n, m, q, s>0$ are singular at $r=0$ and so the model is valid only down to a small finite distance from the origin.

Comparing the $r$-dependences of the terms in Equations
(\ref{eq:mom-r})-(\ref{eq:mom-phi}) we obtain the following relations between the constants:
\begin{eqnarray}\label{eq:q_and_n}
&q&= 2  \, n -2 \, m \, , \\
\label{eq:s_and_n}
&s& = 2 \, n \, .
\end{eqnarray}
In addition, the gravity term in Equations (\ref{eq:mom-r}) and (\ref{eq:mom-theta}) constrains $m=-1/2$. However, if gravity is neglected this constraint disappears and $m$ can take any real value.
Using the ideal gas law (i.e., $p=R \, \rho \, T$) the temperature scaling with $r$ becomes
\begin{equation}
T=\frac{\widetilde{T}(\theta)}{r^{2\, m}} \, ,
\end{equation}
so the temperature increases or decreases with distance from the origin
depending on whether $m<0$ or $m>0$, respectively. 
In the $g\ne 0$ case with $m=-1/2$, $T=r \, \widetilde{T}(\theta)$ and the temperature increases with $r$.
From Equations (\ref{eq:potential-A-r}) to (\ref{eq:potential-psi-theta}) the scaling laws for the potentials are
%
\begin{eqnarray}\label{eq:full_magnetic_potential}
\hat{A}(r,\theta) &=& \frac{\widetilde{A}(\theta)}{r^{n-2}} \, ,\\
\hat{\Psi}(r, \theta) &=& \frac{\widetilde{\Psi}(\theta)}{r^{2 (n-1) -m}} \, ,
\end{eqnarray}
The constancy of $\hat{A}(r,\theta)$ along the field lines implies that
  also $\widetilde{A}/r^{n-2} $ is constant along them. For $n>2$, this implies
that the field line must approach the origin wherever $\widetilde{A}=0$. The poloidal
components of the magnetic field and velocity (Eqs. \ref{eq:potential-A-r} to
\ref{eq:potential-psi-theta}) become
\begin{eqnarray}\label{eq:relationAandB1}
B_r &=& \frac{1}{r^n \sin \theta} \;\widetilde{A}' \, ,\\ \label{eq:relationAandB2}
B_\theta &=& \frac{n-2}{r^n \sin \theta}\; \widetilde{A} \, ,
\end{eqnarray}
and
\begin{eqnarray}\label{eq:vr-function-psi}
v_r &=& \frac{1}{\widetilde{\rho} \, r^{m} \sin \theta}\; \widetilde{\Psi}' \, ,\\ \label{eq:vr-function-psidot}
v_\theta &=& \frac{2(n-1)-m}{\widetilde{\rho} \, r^{m} \sin \theta} \; \widetilde{\Psi} \, ,
\end{eqnarray}
where $()' \equiv d/d \theta$. To simplify the resulting equations, we can also
write the toroidal components in terms of generic functions
$\tilde{b}(\theta)$ and $\tilde{U}(\theta)$ as:
\begin{eqnarray} \label{eq:bphiasfunctionofb}
B_\phi &=& \frac{\tilde{b}}{r^n \sin \theta} \, ,\\ \label{eq:vphiasfunctionofU}
v_\phi &=& \frac{\tilde{U}}{\widetilde{\rho} \, r^{m} \sin \theta} \, .
\end{eqnarray}

We also define a new independent variable, namely,
\begin{equation}
x= 1- \cos \theta \,
\end{equation}
so that $\sin \theta = \sqrt{x\, (2-x)}$ and 
\begin{equation}
\frac{1}{\sin \theta}
\frac{d}{d \theta} = \frac{d}{d x}\, .
\end{equation}
 Finally, the derivatives with respect to
the new angular variable $x$ are indicated with a dot, $\dot{()} \equiv
d /dx$. Using this notation, we see that the magnetic field
components (Equations~\ref{eq:relationAandB1}, \ref{eq:relationAandB2}, and 
\ref{eq:bphiasfunctionofb}) can be written in the form:
\begin{eqnarray} \label{eq:generalconsideration-br}
B_r &=& \frac{\dot{A}}{r^n} \, , \\ 
\label{eq:generalconsideration-btheta}
B_\theta &=& \frac{(n-2)}{\sqrt{x(2-x)}}\, \frac{A}{r^n}\,\\ \label{eq:generalconsideration-bphi}
B_\phi &=& \frac{1}{ \sqrt{x(2-x)}} \,\frac{b}{r^n}\;,
\end{eqnarray}
with corresponding expressions for the components of the momentum in
  terms of $\Psi$, $\dot{\Psi}$ and $U/\rho$. For later use we also include 
here the expression for the electric current:

\begin{eqnarray}\nonumber
 \jvec \;&=& \;
 \frac{1}{r^{n+1}} \left\{ 
\dot{b}\; \er +
\;\frac{n-1}{\sqrt{x(2-x)}}\;b\, \etheta \right. \\
\noalign{\vspace{2mm}}
\label{eq:electric_current}
&-\;& \left. \frac{(n-2)(n-1) A + x (2-x) \ddot{A}}{\sqrt{x(2-x)}
}\;\ephi 
\right\}\,.  \\\nonumber
\end{eqnarray}

With all the previous definitions, the full set of self-similar equations then becomes

%
\begin{widetext}
\begin{eqnarray}
\label{eq:ind-phi}
(2-n) \dot{\Psi} A + \left( 2n-m -2 \right) \dot{A} \Psi &=& 0 \, ,\\ 
\label{eq:ind-pol}
\left( n+m - 1 \right) \left( U \dot{A} -\dot{\Psi} b\right) 
+\rho \, x \left( 2 - x \right) \frac{d}{d x} \left\{ \frac{1}{\rho \, x \left( 2 - x \right) } \left[ \left(2n-m -2\right)\Psi b - (n-2) U A  \right]\right\}&=&0\, ,\\ 
\nonumber
m \, x \left( 2 - x \right) \dot{\Psi}^2-\left( 2n -2 -m \right) x \left( 2 - x \right) \rho \Psi \frac{d}{d x} \left( \frac{\dot{\Psi}}{\rho}\right)+\left( 2n -2-m \right)^2 \Psi^2+U^2+ 2 n \, \rho \, p\, x \left( 2 - x \right) + \\
\label{eq:mov-rad}
+ \rho \left\{ (n-1) \left[ b^2 + (n-2)^2 A^2 \right] + (n-2) x \left( 2 - x \right) A \ddot{A}\right\}-\rho^2  x \left( 1 - x \right)\left( 2 - x \right) \, g &=&0 \, ,\\ 
\nonumber 
 (m-1) (2n-2-m) \Psi \dot{\Psi} -  \left( 2n-2-m \right)^2 \sqrt{x \left( 2 - x \right)} \rho \Psi \frac{d}{d x} \left( \frac{\Psi}{\rho \sqrt{x \left( 2 - x \right)}} \right)+\frac{ \left(1-x \right)}{x \left( 2 - x \right)} U^2 \\ 
\label{eq:mov-the}
-\rho\, \dot{p} \,  x \left( 2 - x \right)- \rho\left[  \, x \left( 2 - x \right)\dot{A}\ddot{A} + b \dot{b}+ (n-2)(n-1)A \dot{A} \right]+\rho^2  x \left( 2 - x \right)\, g  &=&0 \, ,\\ \label{eq:mov-phi} 
2(1-m) \dot{\Psi} U + 2 (2n-2-m) \rho \Psi \frac{d}{dx}\left( \frac{U}{\rho}\right) + 2 \rho \left[ (n-1) \dot{A} b +(2-n) A \dot{b} \right]&=&0\, ,
\end{eqnarray}
\end{widetext}
\noindent keeping in mind that $m=-1/2$ for the particular case when gravity is included
(otherwise it can take on any real value). 
Equations~(\ref{eq:ind-phi})-(\ref{eq:mov-phi}) are derived from 
the poloidal component of the induction equation
(Equation~\ref{eq:induction-poloidal}), the toroidal component of the induction
equation (Equation~\ref{eq:induction-toroidal}), and the three components of the
momentum equation (\ref{eq:mom-r}) to (\ref{eq:mom-phi}). From Equation
(\ref{eq:ind-phi}) we obtain 
\begin{equation}\label{eq:coupling_induction}
|\Psi| = C_\mathrm{ind} |A|^{\frac{2n -m-2}{n-2}} \, ,
\end{equation}
implying that the isosurfaces of $\Psi$ and $A$ coincide, which implies the
fact, already mentioned, that the poloidal components of magnetic field and
velocity are parallel.

\subsection{Behaviour Close to the Vertical
  Axis}\label{sec:generalconsiderations} 

The axisymmetry condition constrains the vector solutions of the
equations. When approaching the vertical axis ($x\to 0$), the
horizontal components of the vector fields must tend to zero,
so the $\theta$ and $\phi$ components must vanish there.
From (\ref{eq:generalconsideration-btheta}), this implies $A(\theta=0)=0$. If
$A$ then behaves as $x^l$ for some constant $l$ when $x\to 0$, we see, from
(\ref{eq:generalconsideration-br}) and
(\ref{eq:generalconsideration-btheta}), that:
%
\begin{eqnarray}
\label{eq:Br_asympt} B_r &\to& x^{\,l-1} \, , \\
\label{eq:Bth_asympt} B_\theta &\to& x^{\,l-1/2} \, .
\end{eqnarray}
Thus, to have a non-singular and non-zero field on the vertical axis ($r>0$, $x=0$), we need to impose $l=1$. Alternatively, if the magnetic field vanishes on the vertical axis, $l>1$. In addition, close to the axis the vertical component is approximately $B_r$ and the horizontal component is approximately $B_\theta$. The resulting inclination $\alpha$ of the magnetic field to the vertical is given by
\begin{equation}
\tan \alpha = \frac{B_\theta}{B_r} \approx \frac{x^{l-1/2}}{x^{l-1}} = x^{1/2} \, ,
\end{equation}
so $\alpha \to 0$ as the vertical axis is approached. This means that the magnetic field becomes vertical when approaching the vertical symmetry axis, as expected.

A similar argument applies to the gas pressure gradient constraining the
shape of $p = \widetilde{p}/r^{2n}$ given by Equations
(\ref{eq:selfsimilardefinitions}) and (\ref{eq:s_and_n}).
The $\theta$-component of the pressure gradient is
\begin{equation}\label{eq:pressure_gradient}
\frac{1}{r} \frac{\partial p}{\partial \theta} = -\frac{1}{r^{2n+1}} \sin \theta \frac{d \widetilde{p}}{d x} = -\frac{\sqrt{x(2-x)}}{r^{2n+1}} \; \dot{\widetilde{p}} \, .
\end{equation}
Assuming the general asymptotic shape $\widetilde{p}\sim
x^{\delta}$ as $x\to 
0$, the $\theta$-component of
the pressure gradient will vanish at the origin only if $\delta>1/2$.
Similarly, the condition that the horizontal components
  vanish at the 
$z$-axis can also be applied to the $v_\theta$, $B_\phi$
and $v_\phi$ components. Assuming that $\rho(x) \sim x^{\sigma}$ with $\sigma
\ge 0$, $\Psi \sim x^{\xi}$, $b\sim x^{\omega}$ and $U\sim x^{\nu}$ and
using Equations (\ref{eq:vr-function-psi})
to (\ref{eq:vphiasfunctionofU}) we find that 
\begin{eqnarray}
\xi -\sigma \ge 1\;, \label{eq:xi_sigma}\\
\noalign{\vspace{2mm}}
\omega > 1/2 \;,\label{eq:omega}\\
\noalign{\vspace{2mm}}
\nu-\sigma \ge1/2 \;. \label{eq:nu_sigma}
\end{eqnarray}
In many situations, the density at the axis
does not vanish implying that $\sigma=0$ and thus $\xi \ge 1$ and $\nu \ge
1/2$.

In this paper we are primarily interested in solutions valid for the half space above $z=0$, hence for the
angular variable $x$ between $0$ and $1$. This allows us to obtain a large family of solutions in which the magnetic field intersects the equatorial
plane. 
Yet, some of the solutions for integer $n$ (positive or negative) allow an extension to the whole space (i.e., $x$ between $0$ and $2$). In those cases, the asymptotic behaviour when approaching the negative $z$ axis must be taken into account, and leads to conditions equivalent to those just discussed [e.g., $A(x=2) = 0$, or Equations~(\ref{eq:Br_asympt}) and (\ref{eq:Bth_asympt}) for $(2-x)$ instead of for $x$]. The condition on $A(x=2)$ guarantees that the field is aligned with the negative vertical axis, but, also, that the total magnetic flux traversing any $r=$const $>0$ spherical surface vanishes. On the other hand, in the cases covering the positive-$z$ semi-space alone, the total magnetic flux traversing a hemi-spherical surface of radius $r$ is $2\,\pi\, r^{2-n} A(x=1)$.

\subsection{Polytropic or Adiabatic Behaviour}\label{sec:polytropic}
Suppose the energy equation can be approximated by a polytropic or adiabatic
law 
\begin{equation}\label{eq:adiabaticequation}
p \, \rho^{-\gamma} = K \, ,
\end{equation}
where $\gamma$ and $K$ are constants.
Then relations 
(\ref{eq:q_and_n}) and (\ref{eq:s_and_n}) 
imply that
\begin{equation}\label{eq:m_f_of_n}
m=\frac{\gamma-1}{\gamma} n \, ,
\end{equation}
or
\begin{equation}
\gamma = \frac{n}{n-m} \, ,
\end{equation}
which implies that the plasma cannot be isothermal (i.e. $\gamma =1$) whenever $m \ne 0$.
Additionally, when gravity is non-zero (hence $m=-1/2$, as seen above), 
\begin{equation}
n=-\frac{\gamma}{2 (\gamma -1)} \, ,
\end{equation}
or equivalently
\begin{equation}
\gamma=\frac{2n}{2n+1} \, .
\end{equation}

\section{The potential and force free field cases}\label{sec:fff_pot}

To obtain the equations for the force-free problem, one must just calculate
the Lorentz force using the self-similar prescriptions for the magnetic field
introduced in Section~\ref{sec:self-similar}, and set it equal to
zero. Equivalently, one can impose the condition that the electric
  current (\ref{eq:electric_current}) be parallel to the magnetic field
  itself
  (\ref{eq:generalconsideration-br})-(\ref{eq:generalconsideration-bphi}). 
That condition yields two independent equations, namely
\begin{eqnarray}                        											\label{eq:fff-rad}
(n-1) \left[ b^2 + (n-2)^2 A^2 \right] + (n-2)\, x \left( 2 - x \right) A\,
  \ddot{A}=0 ~,\\     
%
%
\noalign{\vspace{2mm}}
\label{eq:fff-phi}
(n-1)\, \dot{A}\, b +(2-n) \,A\, \dot{b}=0 ~.
\end{eqnarray}
with the second one, in particular, just stating that the poloidal components
of $\jvec$ and $\Bvec$ are parallel. Integrating the last equation one obtains
\begin{equation}\label{eq:b-of-A}
|b|= c_{\phi} \,{|A|}^\frac{n-1}{n-2}, \quad c_\phi \ge 0\,,
\end{equation}
reflecting the fact that there is no sign relation between $b$ and $A$, i.e.,
that dextral and sinistral twisted structures are equivalent. In the
  general case $n\ne 2$, using (\ref{eq:b-of-A}) in (\ref{eq:fff-rad}) and
  dividing by $A$ one obtains

%
\begin{eqnarray}\nonumber
x \, (2-x) \, \ddot{A} \,&+&\,
 \frac{(n-1)}{(n-2)} \; c_\phi^2 \; \sa \;|A|^{\frac{n}{n-2}} \,+ \\ 
\label{eq:ffe}    &+& \, (n-2) \, (n-1) \, A\quad =\quad 0 \, .
\end{eqnarray}
One can also check that the terms in
Equation~(\ref{eq:ffe}) correspond to the magnetic pressure gradient
(term with the second derivative $\ddot{A}$), the magnetic tension force due
to the curvature of the azimuthal component of the field (term containing
$c_\phi$) and the magnetic tension force associated with the curvature of the
poloidal field (linear term in $A$).

Equation~(\ref{eq:ffe}) was found and studied by \cite{lynden-bell1994}, who assume as boundary conditions that $A=0$ both at $x=0$ and $x=1$,
i.e., both on the vertical axis and on the equatorial plane. This implies
that the latter is necessarily a flux surface. They also focus on solutions
with $A$ positive, which yields field lines that form loops
originating and ending at $r=0$.  They non-dimensionalise
the equation by setting the maximum value of $A$ equal to
unity, and then proceed to solve the second-order differential equation as an
eigenvalue problem to determine the value of $n$ for each value of $c_\phi$.
They find solutions for $n > 2$ using numerical integration.

By considering instead fields in which the field lines start at the origin but go
back down through any part of the equatorial plane and not just the origin,
we explore here more general solutions. This is done by removing the boundary
condition that $A$ vanish on $x=1$, {\bf replacing it by $\dot{A}(0)=1$ and} also allowing for the possibility that $A(x)$ may change sign
within the $(0,1)$ interval. The ODE is then solved as an
initial-value problem. 
In the following, we first consider the potential situation
(Section~\ref{sec:potential-field}) followed by the non-potential but still
force-free case.

\subsection{Potential Field Solutions (${c}_\phi =0$)} \label{sec:potential-field}

From the expression for the electric current
(Equation~\ref{eq:electric_current}), we see that a necessary and sufficient
condition for the self-similar magnetic configuration to be potential is that
\begin{eqnarray}\label{eq:fff-zerobb}
& b=0 \quad \hbox{[equivalently, from (\ref{eq:b-of-A}): $c_\phi =
      0$]}\;,\hfill \\ 
\noalign{\vspace{1mm}
\hbox{and}
\vspace{1mm}}
\label{eq:fff-zerob}
& x \left( 2 - x \right) \ddot{A}+(n-2) (n-1) A=0\;.\hfill
\end{eqnarray}
The potential solution therefore has no azimuthal component. 

\begin{figure}
\begin{center}
\centering\includegraphics[width=0.45\textwidth]{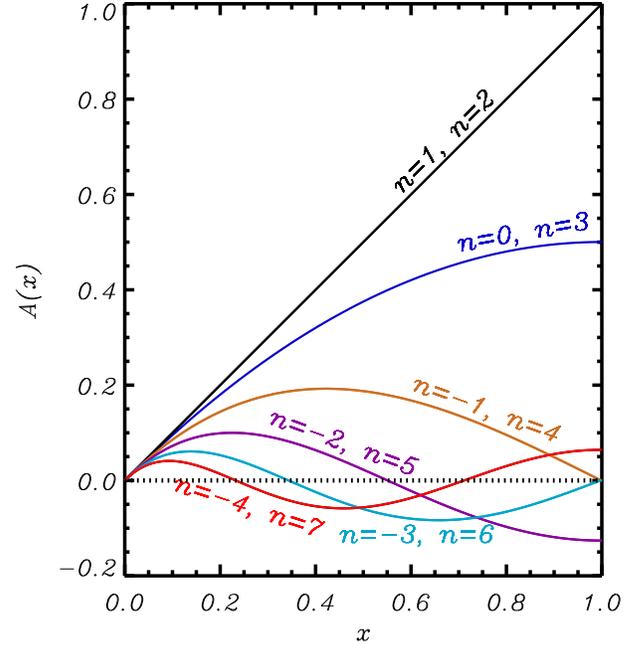}
\caption{Potential  field solutions $A(x)$ for $n= -4$ (or 7), -3 (or 6), -2 (or 5), -1 (or 4), 0 (or 3), 1 (or 2), 2 (or 1), 3 (or 0), 4 (or -1) and 5 (or -2).\label{fig:potentialA}}
\end{center}
\end{figure}

The differential equation (\ref{eq:fff-zerob}) has one solution that is
singular at $x=0$ and one that is non-singular. However, the physical
boundary condition at $x = 0$ (the $z$-axis), namely, that $A(0) = 0$, so
that the horizontal field vanishes, eliminates the singular solution. In
addition, the potential-field equation~(\ref{eq:fff-zerob}) is linear in $A$,
so we can impose the condition $\dot{A}(0)=1$, which implies
$B_r (r,\theta=0) = 1/r^n$, and then obtain the complete
set of solutions through multiplication by an arbitrary constant.
 With these boundary conditions the solution of Equation (\ref{eq:fff-zerob}) is
\begin{equation}\label{eq:sol_potential}
A(x)=\textstyle{\frac{\strut \displaystyle x}{\strut \displaystyle 2}}   \;\,
{}_2^{}F_1^{}\hskip -2pt\left(2-n\,,\,n-1\,;\,2\,;\, \textstyle{\frac{1}{2}} x\right)~, 
\end{equation}
in terms of the hypergeometric function $_2F_1$ \citep[see][]{abramowitz1972}. It is plotted in Figure~\ref{fig:potentialA} for different values of the parameter $n$. We see that the solutions for $n$ and $3-n$ coincide, which follows from the
dependence of Equation~(\ref{eq:fff-zerob}) on $n$ exclusively through the quadratic factor $(n-2)(n-1)$. The corresponding magnetic field lines are different, however, since $B_r$ and $B_\theta$ have additional dependences on
$n$ through their variation with $r$, see Equations~(\ref{eq:generalconsideration-br}) and (\ref{eq:generalconsideration-btheta}).

\begin{figure}
\begin{center}
\centering\includegraphics[width=0.24\textwidth]{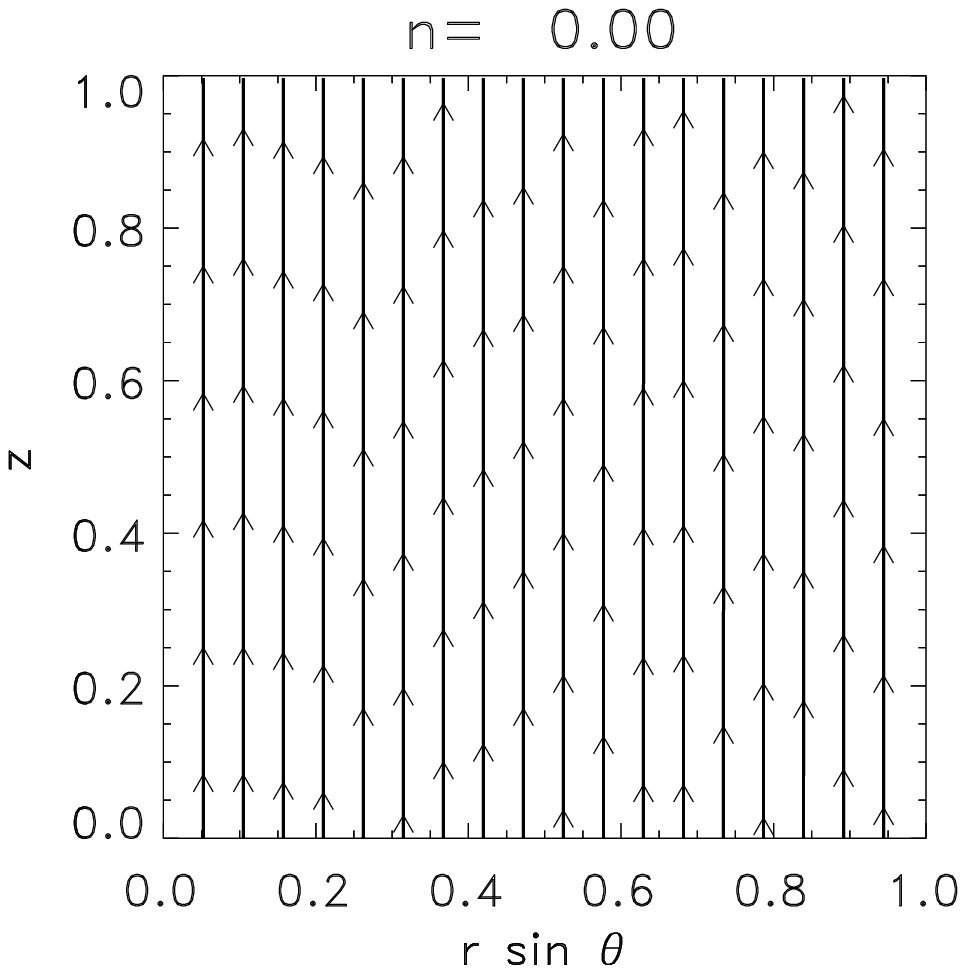}\includegraphics[width=0.24\textwidth]{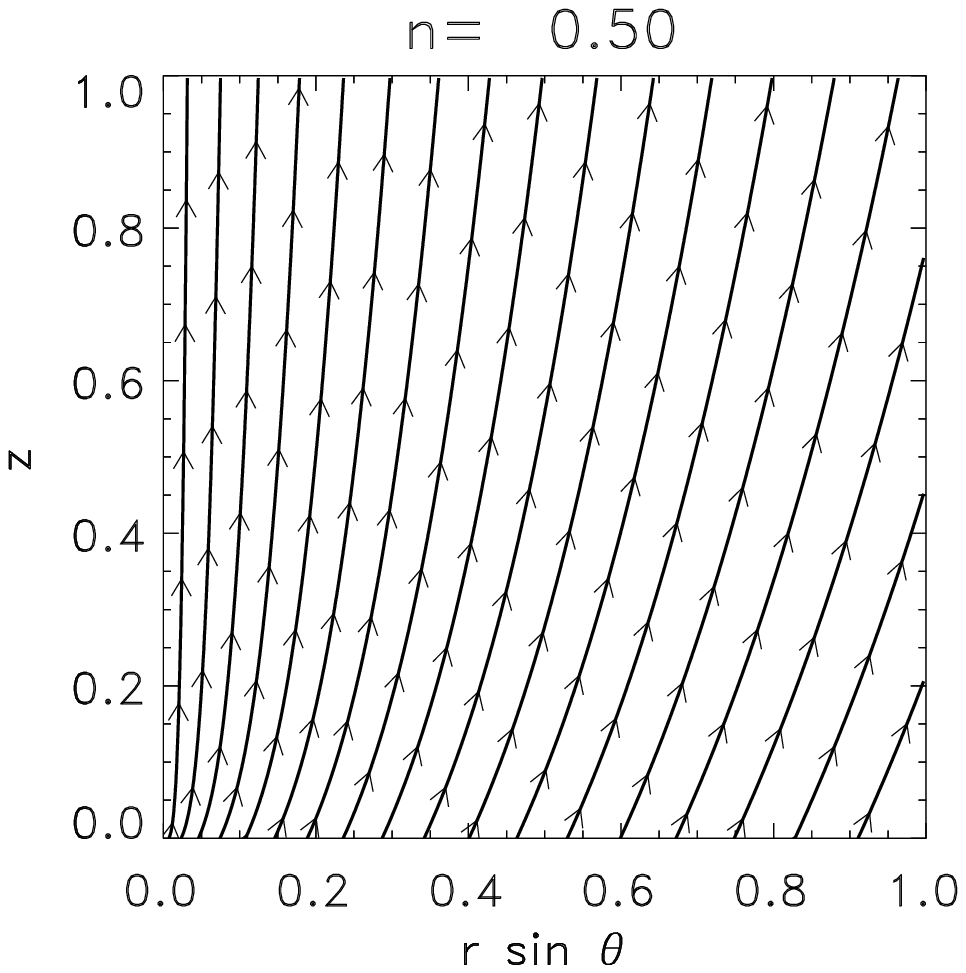}\\
\centering\includegraphics[width=0.24\textwidth]{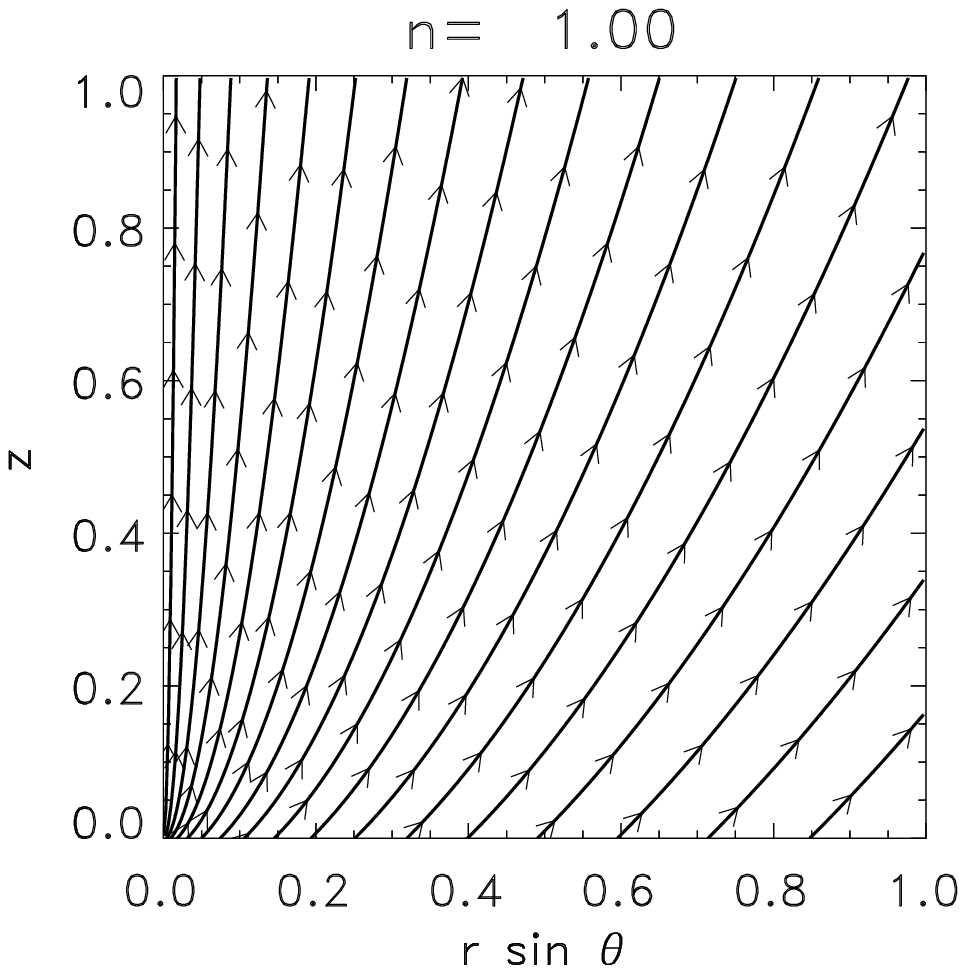}\includegraphics[width=0.24\textwidth]{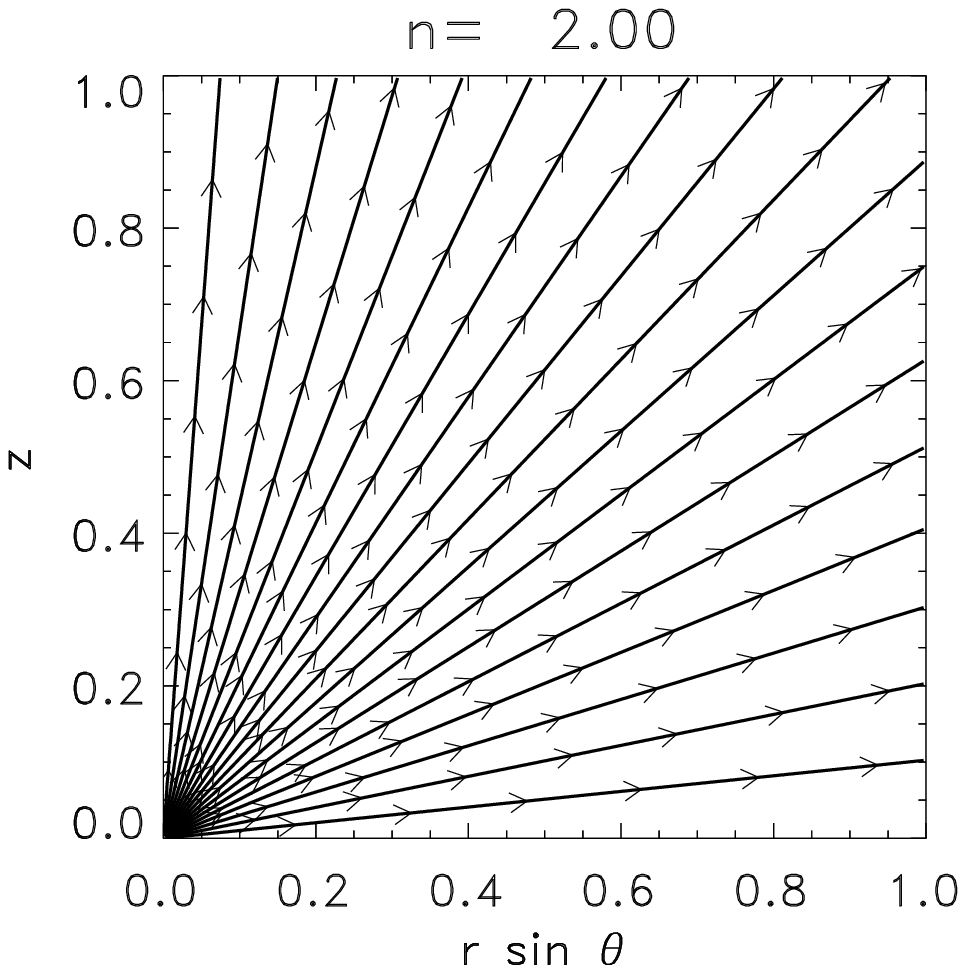}
\end{center}
\caption{Potential field solutions, showing the poloidal magnetic field for $n=0, 0.5, 1$, and $2$.\label{fig:poloidalfieldpotential1}}
\end{figure}

\begin{figure}
\begin{center}
\centering\includegraphics[width=0.24\textwidth]{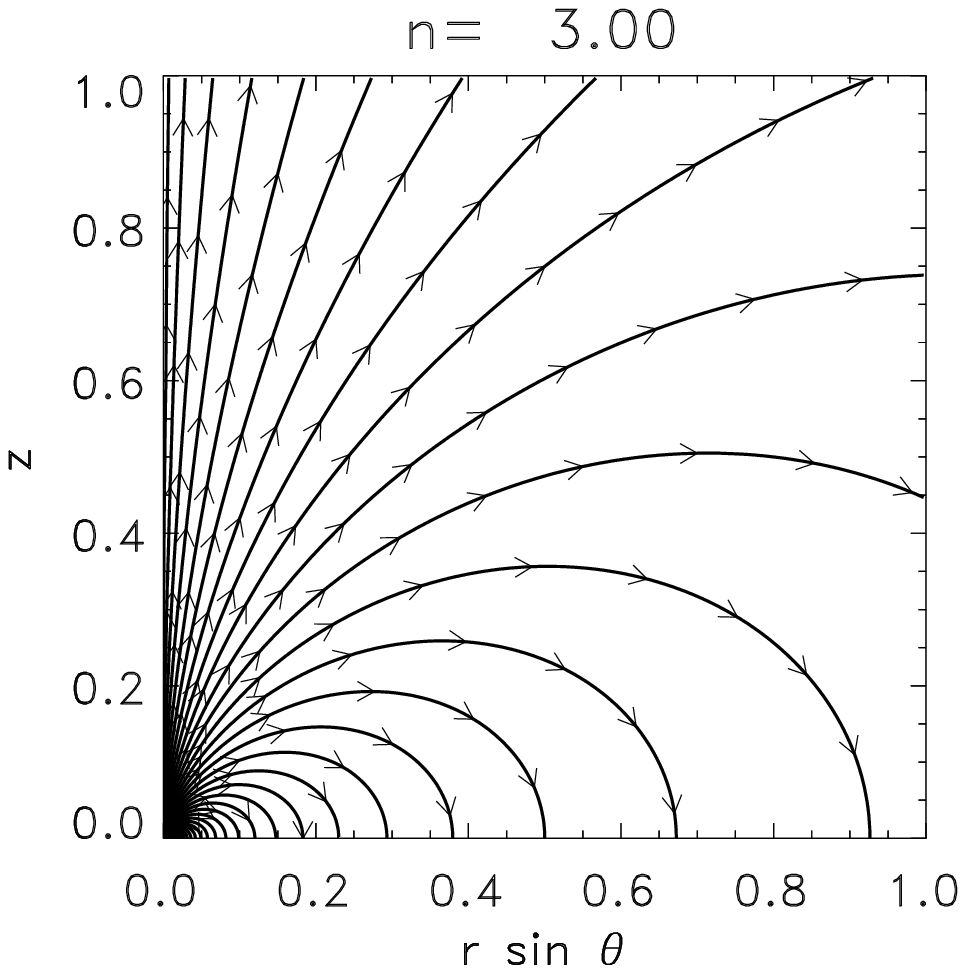}\includegraphics[width=0.24\textwidth]{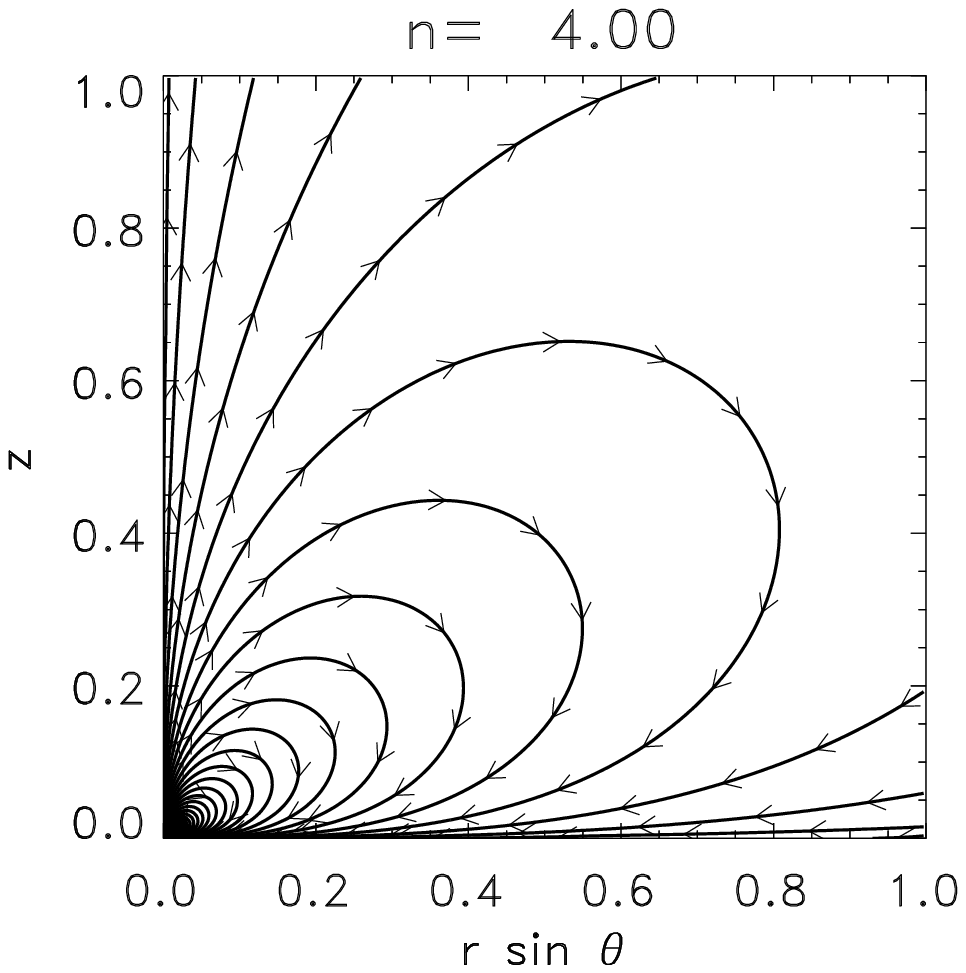}\\
\centering\includegraphics[width=0.24\textwidth]{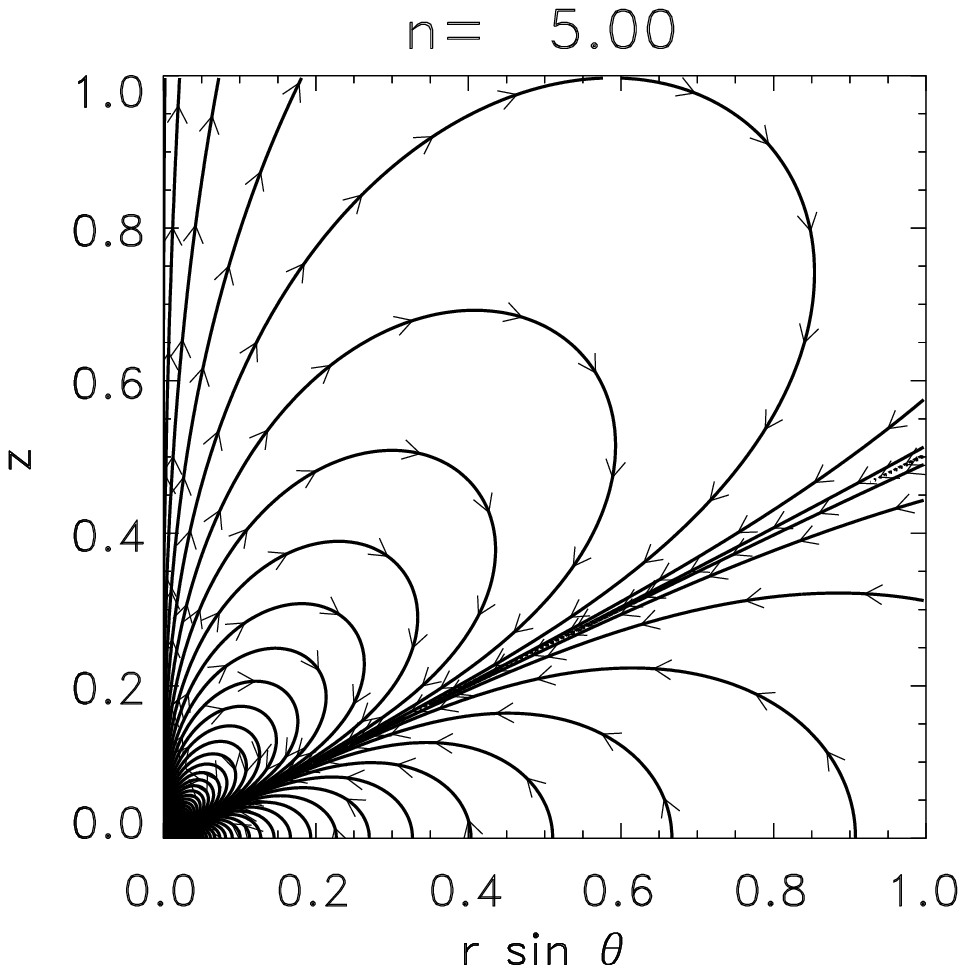}\includegraphics[width=0.24\textwidth]{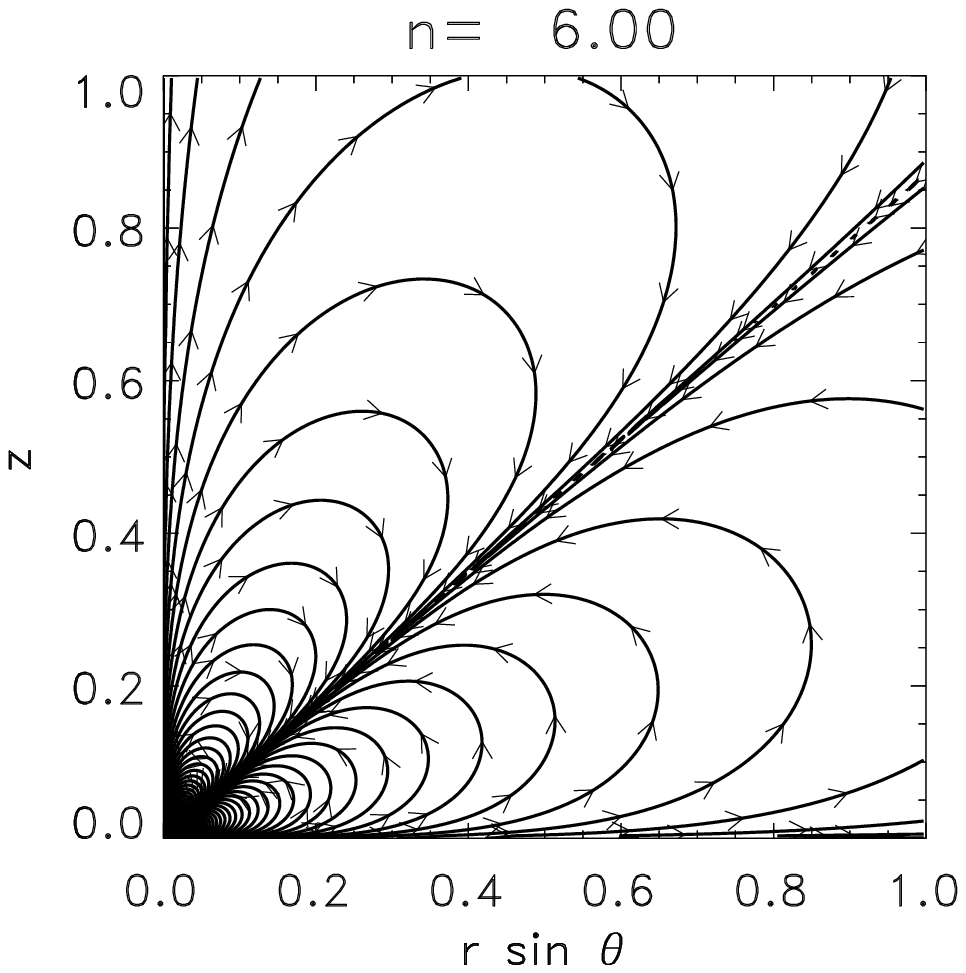}\\
\centering\includegraphics[width=0.24\textwidth]{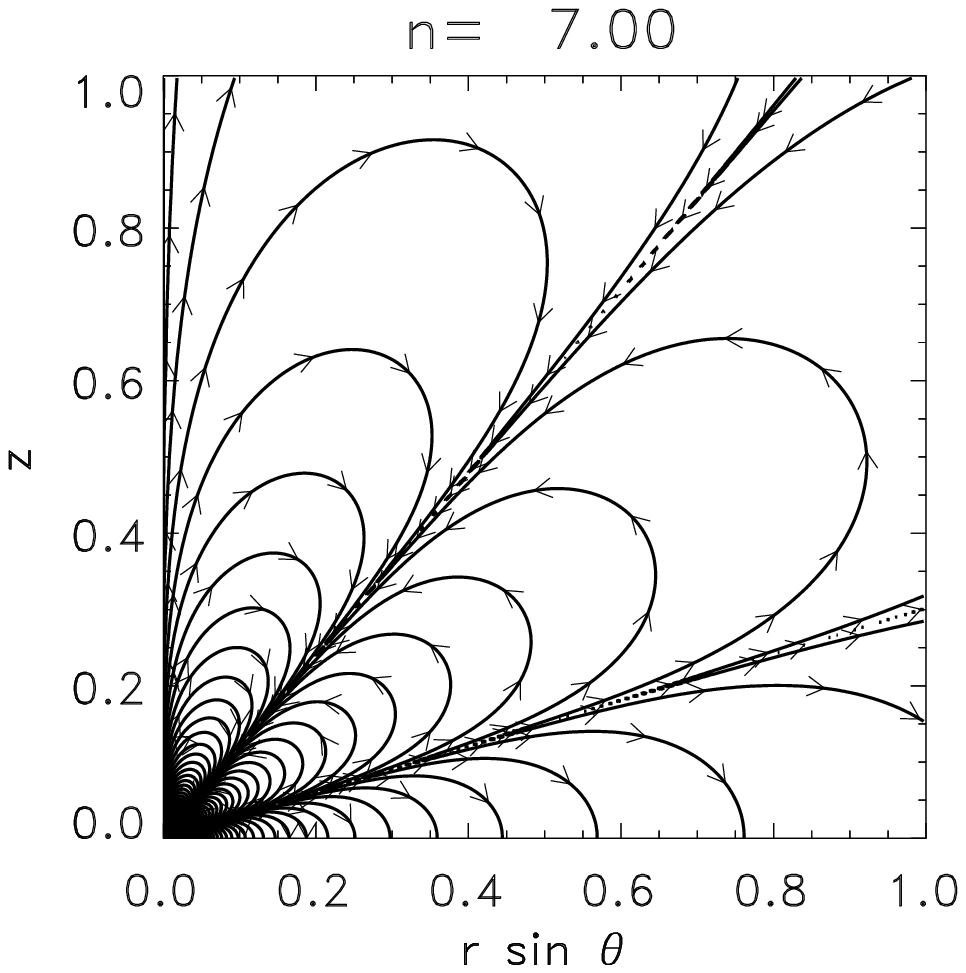}\includegraphics[width=0.24\textwidth]{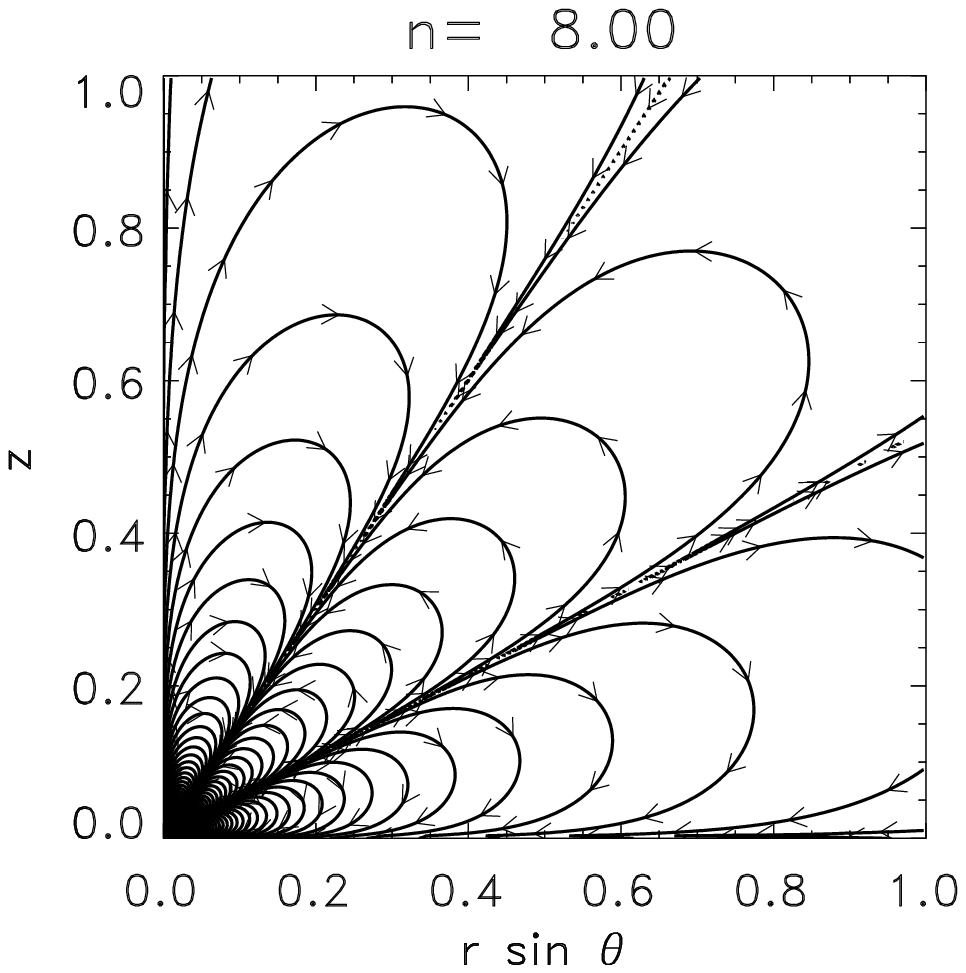}
\end{center}
\caption{Potential field solutions, showing the poloidal magnetic field for $n=3, 4, 5, 6, 7$, and $8$.
}\label{fig:poloidalfieldpotential2}
\end{figure}

\begin{figure}
\begin{center}
\centering\includegraphics[width=0.24\textwidth]{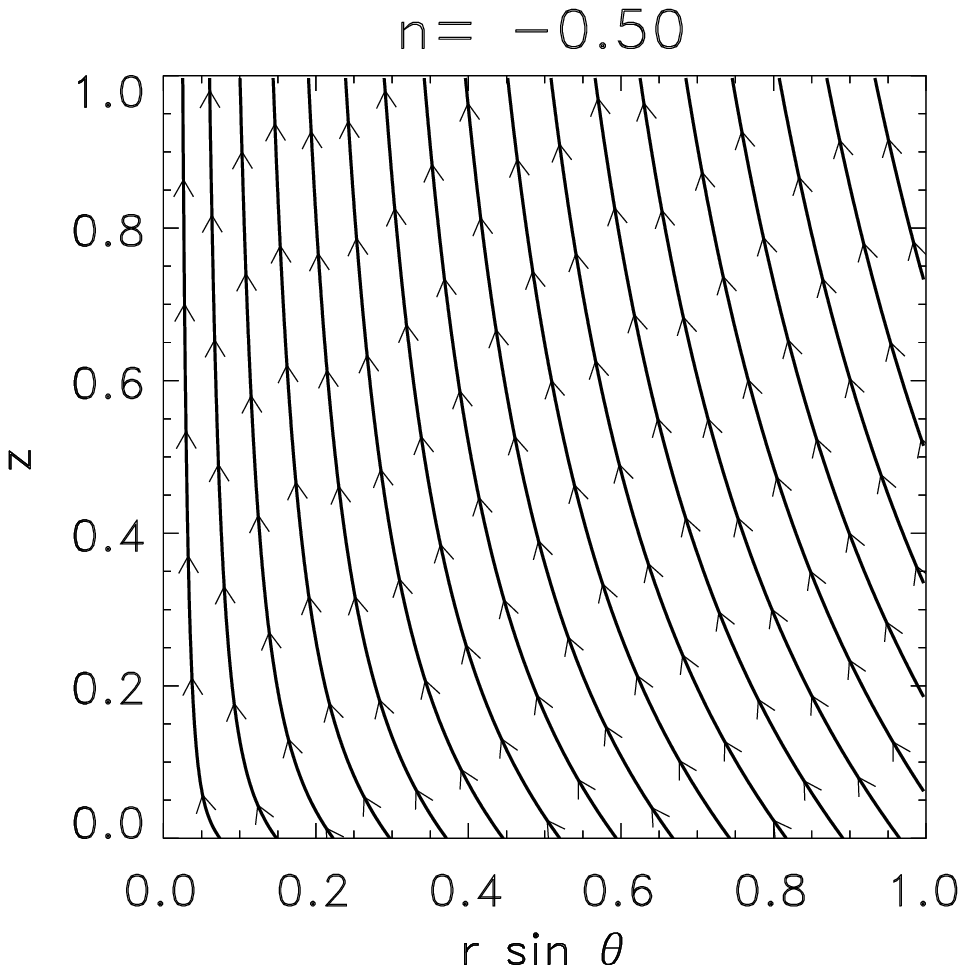}\includegraphics[width=0.24\textwidth]{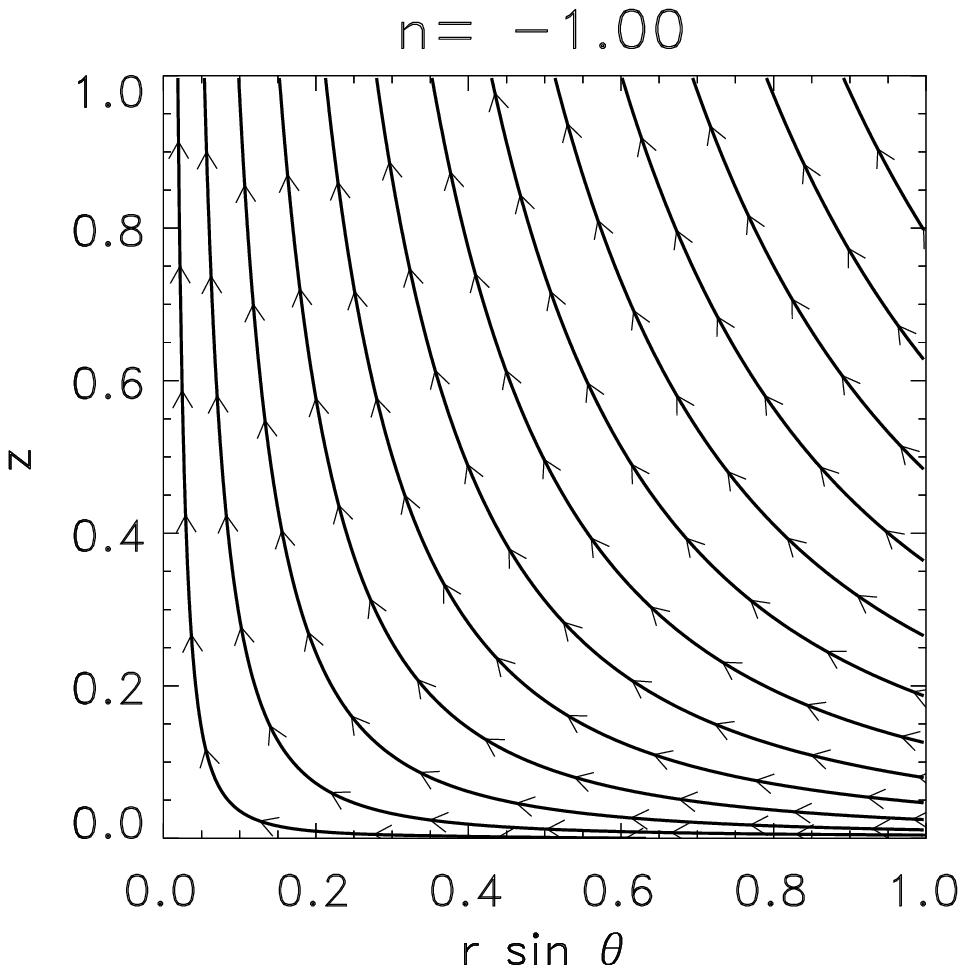}\\
\centering\includegraphics[width=0.24\textwidth]{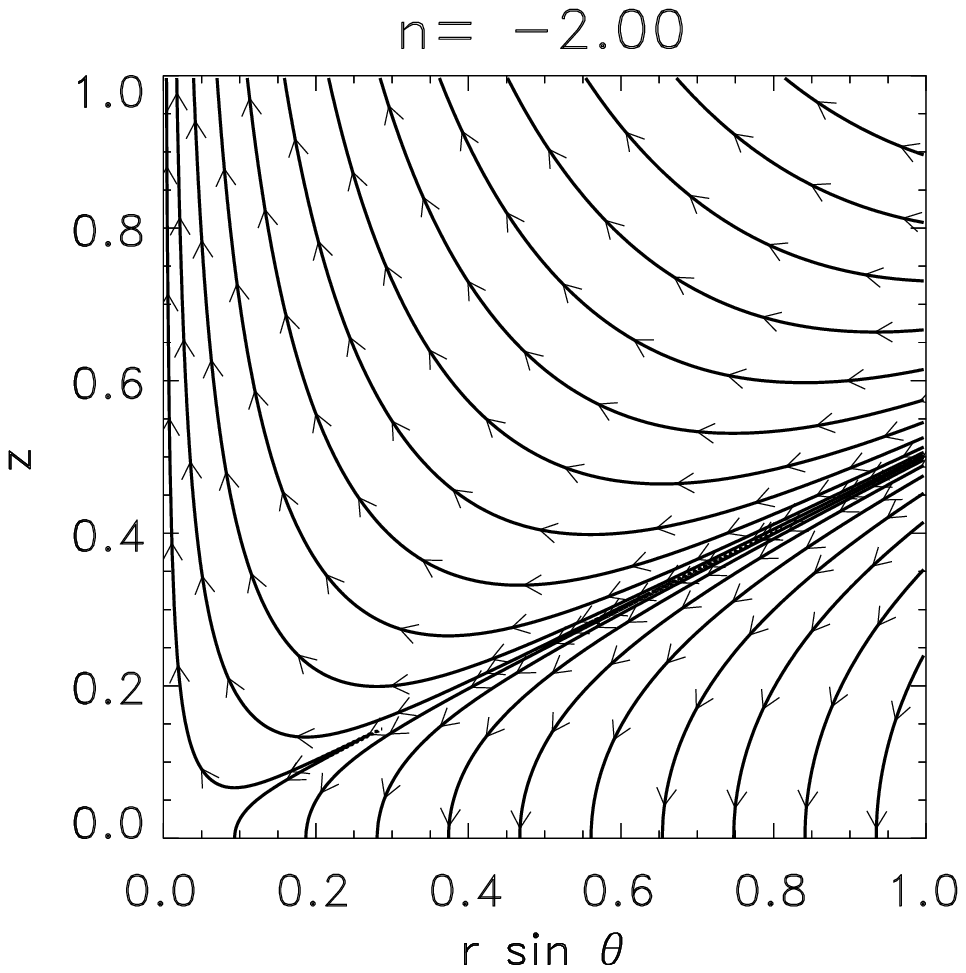}\includegraphics[width=0.24\textwidth]{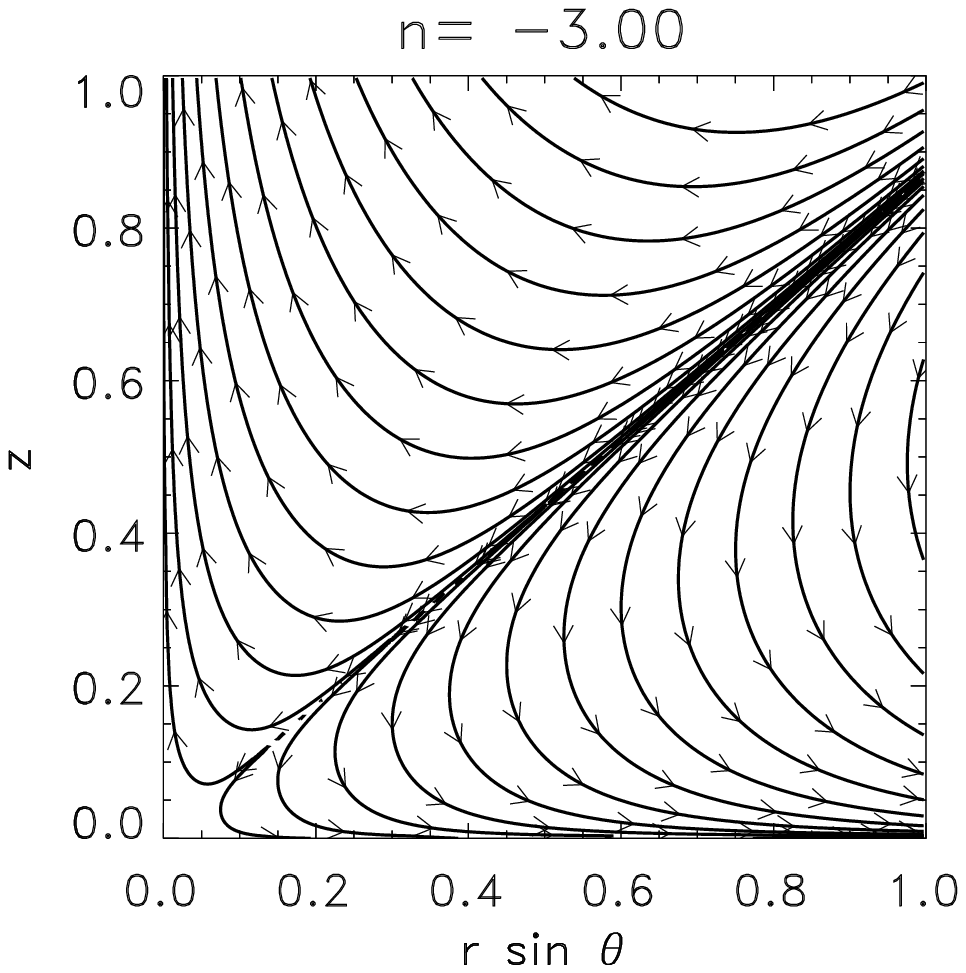}\\
\centering\includegraphics[width=0.24\textwidth]{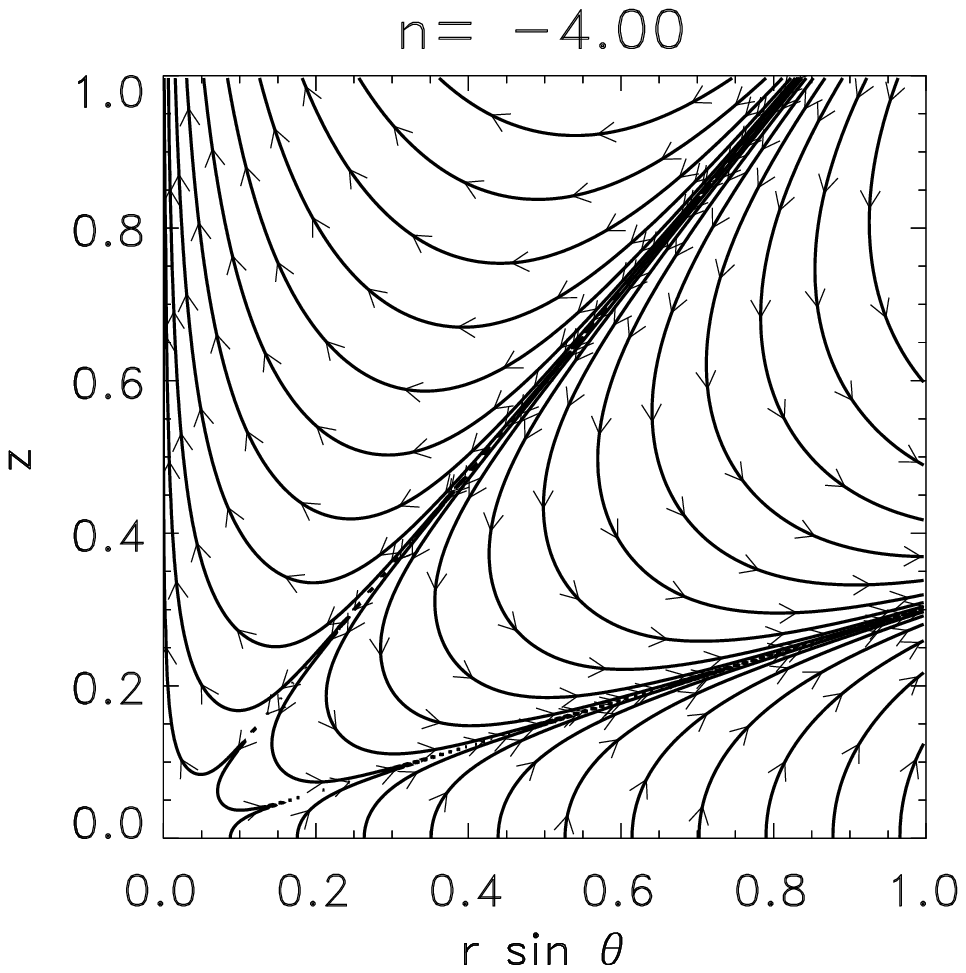}\includegraphics[width=0.24\textwidth]{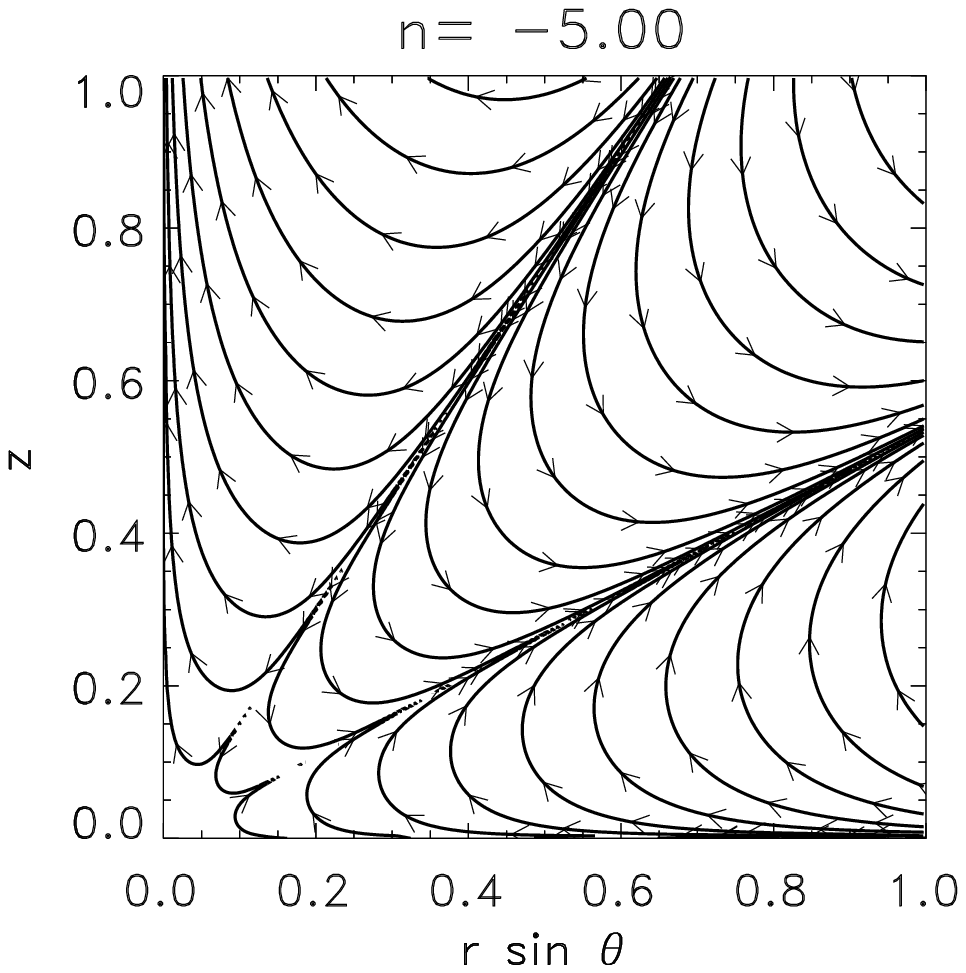}
\end{center}
\caption{Potential field solutions, showing the poloidal magnetic field for $n=-0.5, -1, -2, -3, -4$ and $-5$.}\label{fig:poloidalfieldpotential3}
\end{figure}

To visualize the magnetic field lines, we use the fact, explained in
  Section~\ref{sec:basic_equations} following
  Equation~(\ref{eq:magnetic_potential}), that the isolines of the magnetic
  potential $\hat{A} = A(x)/ r^{n-2}$ in the poloidal plane are poloidal
  field lines. Further, equally spaced values of $\hat{A}$ correspond to
isolines that contain the same amount of poloidal flux between them
\citep[see Sect. 2.9.3 of][]{priest2014}. For the diagrams in this section
(Figures~\ref{fig:poloidalfieldpotential1},
\ref{fig:poloidalfieldpotential2}, and \ref{fig:poloidalfieldpotential3}),
however, for better visualization, we are choosing non-equally spaced values
of $\hat{A}$ for the isolines.
On the other hand, the arrow heads on the field lines in those figures
indicate the direction of the magnetic field.

In Figures~\ref{fig:poloidalfieldpotential1},
\ref{fig:poloidalfieldpotential2}, and \ref{fig:poloidalfieldpotential3}, we
have plotted the poloidal field lines for various cases with $n$ in the
ranges $0 < n \le 2$, $n>2$, and $n<0$, respectively. For $n>0$ the origin is
a singularity and the field strength declines with distance along each
radius. In contrast, for $n<0$ the origin is a null point and the magnetic
field strength increases with distance from the origin. For $n=0$ the
magnetic field is uniform in space and parallel to the $z$-axis (see
Fig. \ref{fig:poloidalfieldpotential1}a). For $0 < n < 2$ the field lines
bend towards the $z$-axis as $z$ increases
(Figure~\ref{fig:poloidalfieldpotential1}, panels b and c).  For $n=2$
  equation (\ref{eq:fff-zerob}) strongly simplifies to just $\ddot{A} = 0$. The
  field lines in this case are straight, radiating in all directions from the
  origin (Figure~\ref{fig:poloidalfieldpotential1}d) -- this solution
  corresponds to a pure magnetic monopole located at the origin of
  coordinates.

For $n > 2$ the field lines curve away from the $z$-axis as $z$ increases, and form loops, closing down to meet the horizontal axis
(Fig. \ref{fig:poloidalfieldpotential2}c). For $n=3$ the field lines are
vertical where they meet the surface $z=0$. For $n=4$ the field lines are
loops that start and close at the origin with one lobe in between. For $n=5$
the structure is more complex with one lobe and one magnetic separatrix. In
this situation the magnetic field is also vertical at the surface $z=0$. For
larger $n$ values more complex structures are obtained such that for $n=6$
there are 2 lobes, for $n=8$ there are 3 lobes, for $n=10$ there are 4 lobes
and so on. 

 In Figure \ref{fig:poloidalfieldpotential3}, we have plotted several cases
 with $n<0$, for which the magnetic field vanishes at the origin. The
 solutions with integral negative $n$ are particularly simple: they have field distributions on the horizontal plane that are either horizontal
 or vertical and the solution can easily be extended to the whole space
 including negative values of $z$. The case $n=-1$ turns out to be a standard
 3D null point with the $z$-axis being the spine and the field in the
 horizontal plane being the fan \citep{priest1996}. For arbitrary negative
 (but still integer) $n$, the value of $-n$ determines the number of fan
 surfaces (fan cones, in fact) possessed by the solution: $n=-5$, for
 instance has five fan surfaces (of which three are apparent in the quadrant
 shown in Figure~\ref{fig:poloidalfieldpotential3}, bottom right panel).
In the general case when $n$ is not an integer, the magnetic field intersects
the horizontal axis at an angle different from $0$ or $\pi/2$: for example,
the ranges $3<n<4$ or $-2<n<-1$ yield an inclination between $\pi/2$ and $0$,
while for $2<n<3$ or $-1<n<0$ the inclination lies between $\pi/2$ and $\pi$.

Simple trigonometric expressions can be found as solutions for $A$ in
Equation~(\ref{eq:fff-zerob}) when $n$ is a positive or negative integer. Those
solutions indicate the number of lobes (for $n>0$) or fan
surfaces (for $n<0$) found above, and also justify the fact that the
corresponding field lines are either tangential or perpendicular to the
equatorial plane. We recall that the solution for $A$ is identical for $n$
and $3-n$. These trigonometric expressions can be found using the relation between the Hypergeometric functions, $_2F_1$ and the associated Legendre polynomials, $P^{1}_{n-2}(1-x)$.
When $n=-1$ or $4$, $A= \cos \theta \sin^2 \theta$, such that $n=4$ gives a
single lobe and $n=-1$ gives a first-order null point with one spine and one
fan surface. On the other hand $n=-3$ or 6 gives $A=\cos \theta \sin^2 \theta
(3 - 7 \cos^2 \theta )$, such that $n=6$ has two lobes and $n=-3$ gives a
third-order null with two fans (see
Fig. \ref{fig:poloidalfieldpotential3}). Again $n=8$ gives three lobes, while
$n=-5$ has three fans, and so on.

\subsection{Non-potential Force-Free Field Solutions 
(${c}_\phi \ne 0$)}
\label{sec:fff}

To obtain solutions for the non-potential force-free case, one must
  solve Equations~(\ref{eq:b-of-A}) and (\ref{eq:ffe}) now with $c_\phi \ne 0$. 
  We note in passing that the force-free $\alphaff$ coefficient such that
  $\nabla \times   \Bvec = \alphaff\,\Bvec$ can be written in our case:
\begin{equation}\label{eq:alpha_ff}
\alphaff = \pm\, \frac{n-1}{n-2} \,c_\phi\;\frac{|A|^{\frac{1}{n-2}}}{r}\;
  \;=\; \, \pm \,
  \frac{n-1}{n-2} \frac{|b|^{\frac{1}{n-1}}}{r} \,.
\end{equation}
\noindent The first expression says that $\alphaff$ is essentially
  $c_\phi$ times a power of the magnetic potential $\hat{A}$ whose isolines
  coincided with the poloidal field lines: this agrees with
  the fact that $\alphaff$ must be invariant along each field line.  From
  (\ref{eq:alpha_ff}) we also see that the $c_\phi$ parameter provides a
  measure of the field line twist.  

Equation (\ref{eq:ffe}) is non-linear in $A$, yet it fulfills the following
scaling law: given any arbitrary constant $\lambda >0 $, if $A=f(x;c_\phi)$
is a solution, then
\begin{equation}\label{eq:scaling_law}
A=\lambda f(x;c_\phi \lambda^\frac{1}{n-2})
\end{equation}
is also a solution. So, once we have found a solution for a given $c_\phi$,
we immediately have a whole family of equivalent solutions for all strictly
positive values of $c_\phi$. Alternatively: imagine you impose constraint
$\dot{A}(0)=1$ and find solutions of the equation for all admissible
$c_\phi$. Then you immediately have solutions for all other values of
$\dot{A}(0)$ just by using the scaling law (\ref{eq:scaling_law}). In this
sense we will use here again the normalization $\dot{A}(0)=1$ of Section
\ref{sec:potential-field} without loss of generality. Additionally, as
  explained earlier, one must impose the boundary condition that $A(0)=0$,
  for the field to be vertical on the $z$-axis.
There exist solutions of Equation (\ref{eq:ffe}) only for $n> 2$. For $n<2$ 
it is not possible to find a solution, since $|A|^\frac{n}{n-2}$ is singular
and cannot be compensated by the other terms on the
left-hand side: on the one hand, $A(x \to 0) \to 0$; on the other, 
the term $x(2-x) \ddot{A}$ behaves as $x^{l-1}$ when $x \to 0$, so it also
tends to $0$ since $l \ge 1$ as seen above to make the horizontal field components vanish at the axis.
We conclude that $n<2$ is a forbidden region for self-similar force-free
field solutions. In particular, there  are no solutions with $n<0$.

\begin{figure}
\includegraphics[width=0.45\textwidth]{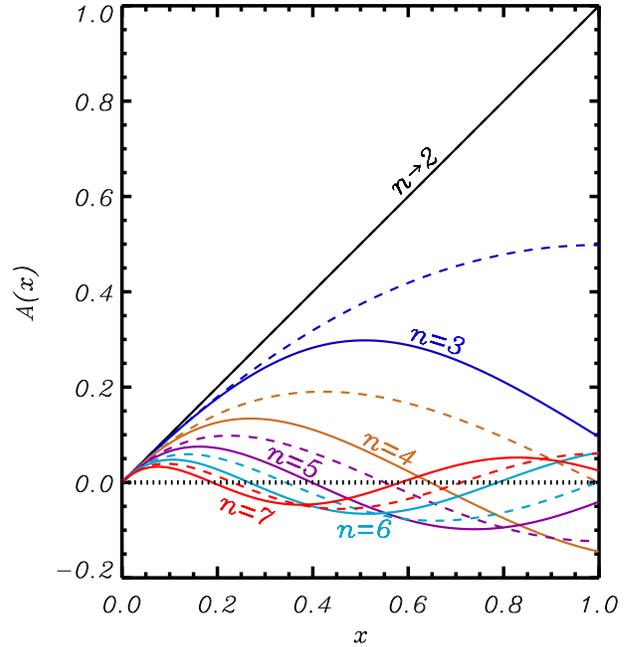}
\caption{Force-free solutions $A(x)$ for different values of $n\ge2$. The solid lines correspond to the force-free field solutions with $c_\phi=5$. The dashed lines are the equivalent potential solutions (Sec. \ref{sec:potential-field}).}\label{fig:fffA}
\end{figure}

\begin{figure}
\begin{center}
\includegraphics[width=0.24\textwidth]{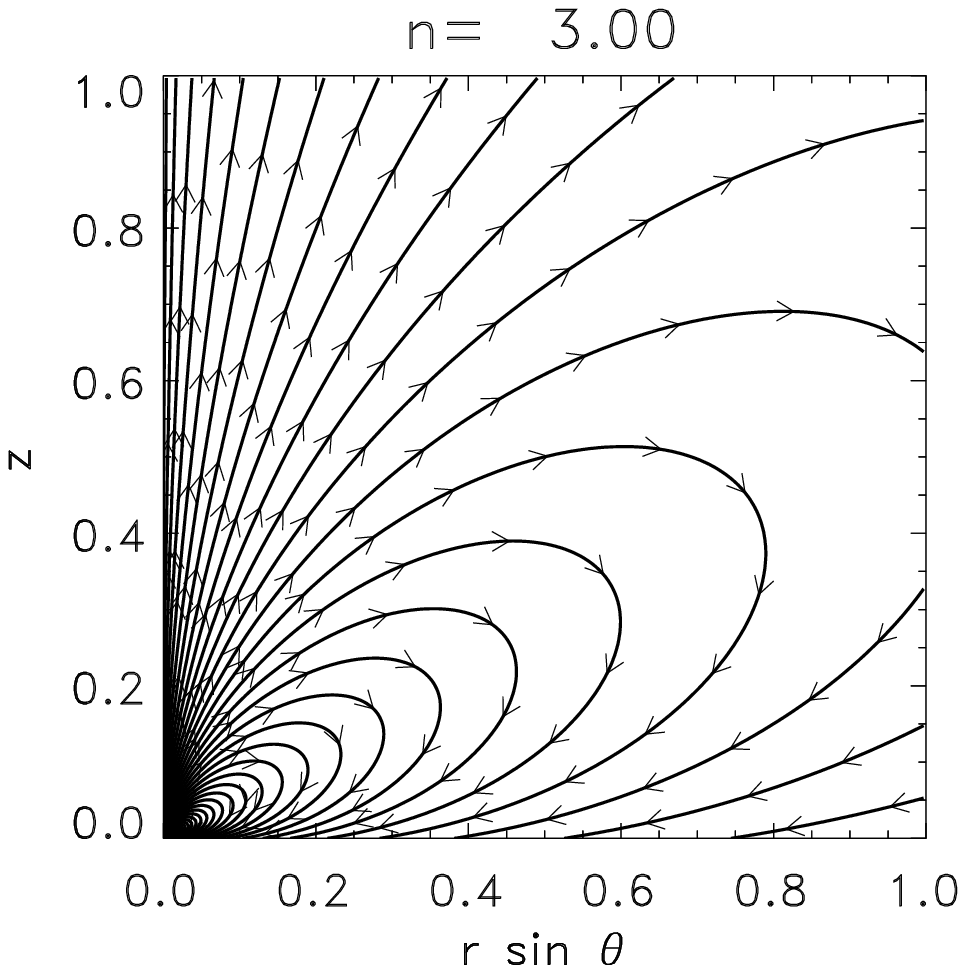}\includegraphics[width=0.24\textwidth]{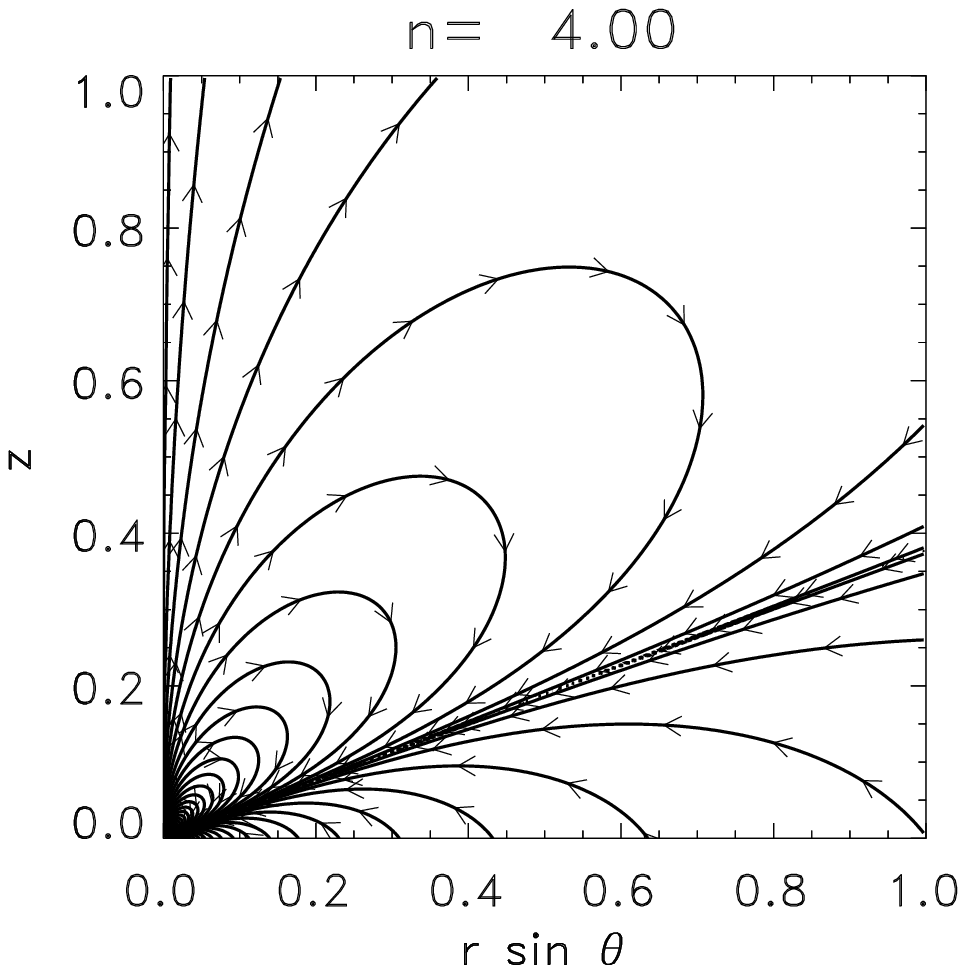}\\
\includegraphics[width=0.24\textwidth]{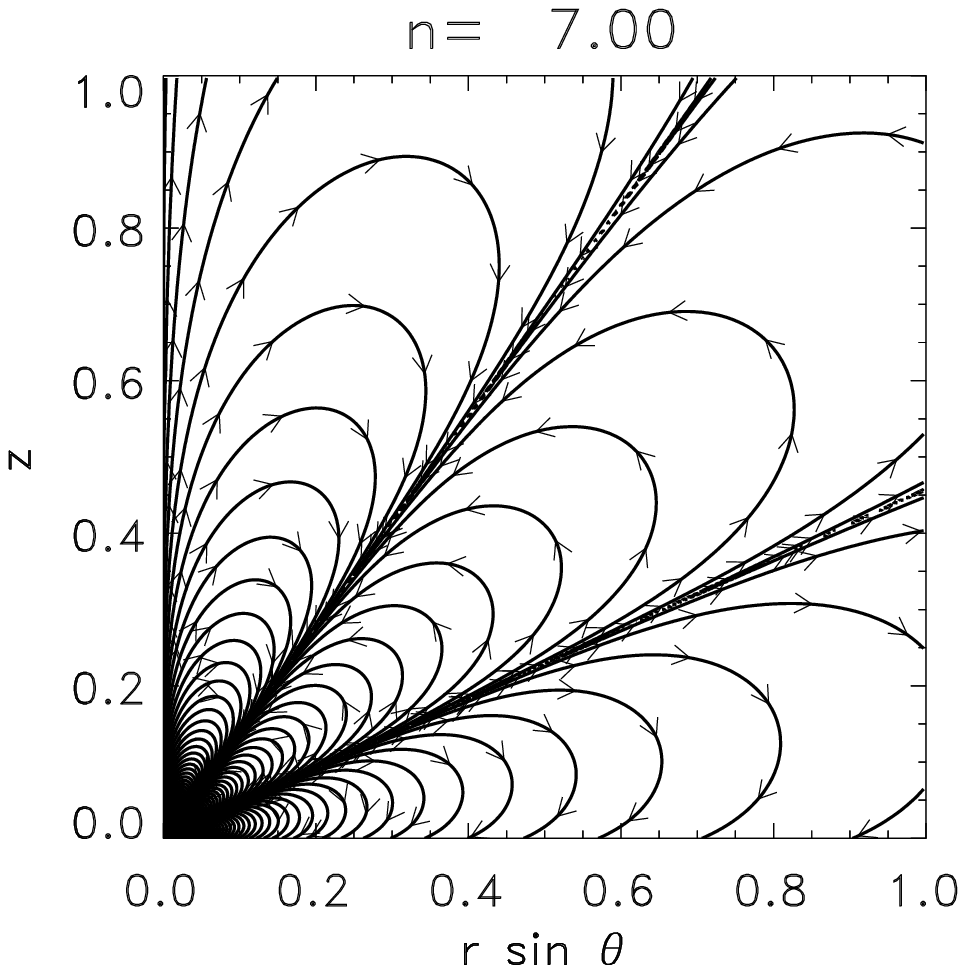}\includegraphics[width=0.24\textwidth]{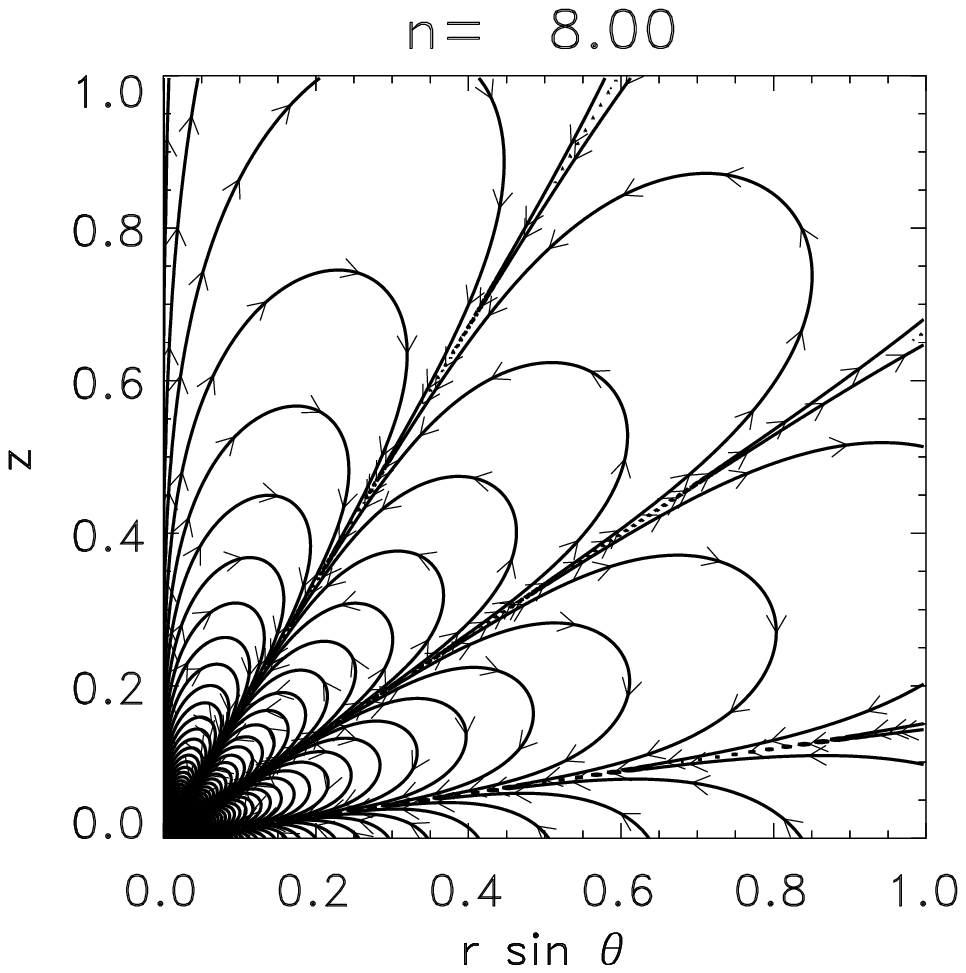}
\end{center}
\caption{Force-free field solutions, showing the poloidal magnetic field for $n=3, 4, 7$ and 8 when $c_\phi=5$.
}\label{fig:poloidalfield-fff1}
\end{figure}

We have solved numerically Equation (\ref{eq:ffe}) for $n>2$
  with the boundary conditions mentioned above: the result, for the
  particular case $c_{\phi} = 5$,  can be seen in
  Figure~\ref{fig:fffA}. To highlight the difference to the potential
  solutions, the latter have been overplotted as dashed lines. 
  In general the non-potential solutions have their extrema closer to the
  $x=0$ axis than the potential ones. For example, for $n=4$ the
    maximum of $A(x)$ is located at $x=0.4$ in the potential case but at
    $x=0.25$ in the force-free case. Also: the solution crosses the
    horizontal axis at $x=1$ for the potential case but at $x=0.63$ for the
    force-free situation. For a sufficiently large $c_{\phi}$ additional
    extrema and crossing points will appear between $x=0$ and $x=1$ in the force-free solutions when compared with the potential cases.
The curve for $n \to 2$ in Figure~\ref{fig:fffA} has been obtained as
an asymptotic limit: equation~(\ref{eq:ffe}) becomes singular in that limit,
but the solutions for $A(x)$ tend to the straight line shown for values of $n$
increasingly close to $2$. This limit has been studied by
\citet{lynden-bell1994}: they find a singularity at a surface
$\theta=$const (called  \textit{sheet discharge} by them) 
which contains purely azimuthal field while in the remaining volume 
the field is purely radial.

\begin{figure}
\begin{center}
\includegraphics[width=0.5\textwidth]{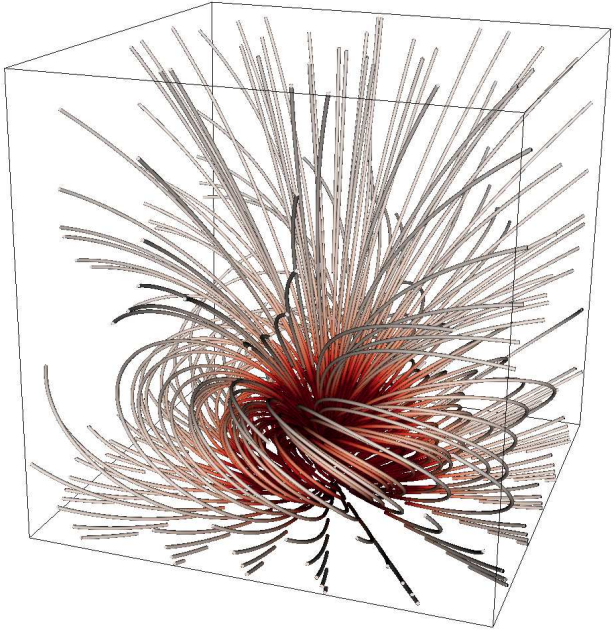}\\
\end{center}
\caption{Three-dimensional plot of the force-free magnetic field lines for $n=3$ and 
$c_\phi=5$. The red color of the lines indicates the intensity of the magnetic field. The darkest red indicate the stronger field intensity and the grey indicates the weaker values.\label{fig:3dfff-1}}
\end{figure}

In Figure \ref{fig:poloidalfield-fff1} the poloidal field lines are plotted
for $n>2$. There are important differences compared to the
potential situation without twist in Figure
\ref{fig:poloidalfieldpotential2}. For example, the potential field with
$n=3$ is perpendicular to the horizontal axis; however, when twist is added
the field lines are considerably inclined with respect to
the vertical direction. In fact in this situation the field
  lines are almost closed. In general, the field lines are
 modified to a large extent at the bottom of the figures with respect
to the potential field situation. There is a correspondence between the fans
and lobes in both situations for all $n$ shown in the figure.

A balance between the magnetic pressure gradient and poloidal tension non
longer holds and so the poloidal field is modified to compensate for the
azimuthal magnetic tension.  In Figure \ref{fig:3dfff-1}, the
three-dimensional field lines for $n=3$ are shown. 
The magnetic field forms loops starting at the centre and returning to the bottom surface. The magnetic field intensity decreases with distance from the centre of the structure. The field structure has twist that is more pronounced close to the horizontal axis in this particular case. The
reason is that $b_\phi$ and so the twist depend on $A(x)$
(Equation~\ref{eq:b-of-A}). This function has a maximum close to $x=0.5$ as
we see in Figure \ref{fig:fffA} which corresponds to an angle of
approximately 60 degrees with respect to the $z$-axis.

\section{Non force-free solutions without poloidal flows
  ($\Psi=0$)}\label{sec:pure-rotation} 

In this section we consider a more general case in which
  the Lorentz force is allowed to be non-zero. For the sake of compactness
  only specific cases are considered that can be analyzed in some detail. We
  focus on problems without poloidal flow by setting $\Psi=0$ but consider
  non-zero values of the density $\rho$. 
Equation 
(\ref{eq:ind-phi}) 
is then trivially satisfied. The
$\phi$-components of the induction and momentum equations, (\ref{eq:ind-pol})
and (\ref{eq:mov-phi}), become:
\begin{eqnarray}\nonumber
\left(n+m-1 \right) \frac{U}{\rho\,x\, \left(2-x\right)}\, \dot{A} ~~~~~~~~~~~~~~~~~~~~&&\\  \label{eq:ind-pol_case3}
\,-\,\frac{d}{d x} \left\{ (n-2) \frac{U}{\rho \, x \left( 2 - x \right)} \,A \right\} \; &=&0\; , \\ \label{eq:mov-phi_case3}  
(n-1)  \dot{A} \,b +(2-n) A \,\dot{b}   \; &=&0 \; .
\end{eqnarray}
These equations have simple algebraic solutions: 
\begin{eqnarray}
\label{eq:u-purerotation_2} {|U|\over \rho}\,&=&\, c_U\, \, x \,\left( 2 - x
  \right) |A|^\frac{(m + 1)}{(n - 2)} \, ,\\ 
\label{eq:b-purerotation} 
|b|&=& c_\phi\;\; {|A|}^{\frac{n-1}{n-2}} \, , 
\end{eqnarray}
\noindent where the last equation is identical to Equation (\ref{eq:b-of-A})
for the force-free field situation: in the absence of poloidal flows, the
$\phi$-component of the Lorentz force must vanish, which is the condition
that led to Equations (\ref{eq:fff-phi}) and (\ref{eq:b-of-A}). The
integration constants $c_U$ and $c_\phi$ are positive.  
The two remaining equations in the system
(\ref{eq:ind-pol})--(\ref{eq:mov-phi}) are the radial and $\theta$ components
of the momentum equation, which for $\Psi=0$ simplify to:
\begin{eqnarray} \label{eq:fundamental1}
&& \frac{U^2}{\rho}+ 2 n \, \, p \, x \left( 2 - x \right) +\\ \nonumber
&& ~~~~~~~~~~~ -\rho g x \left( 1 - x \right)\left( 2 - x \right)\,   + A \, \mathcal{F}(A,\ddot{A})=0 \, , \\ \label{eq:fundamental2}
&& \frac{\left(1-x  \right)}{x \left( 2 - x \right)} \frac{U^2}{\rho} - \dot{p} \, x \left( 2 - x \right)  +\\ \nonumber
&& ~~~~~~~~~~~ + \rho \,g\, x\, \left( 2 - x \right)\, - \frac{\dot{A}}{n-2}\mathcal{F}(A,\ddot{A})=0 \, ,
\end{eqnarray}
where
\begin{eqnarray}\label{eq:F_definition}
\mathcal{F}(A,\ddot{A}) &\equiv & (n-1)\; \sa \; c_\phi^2\; |A|^{n \over
  n-2} 
+ \\
\noalign{\vspace{1mm}}
\nonumber
&&(n-2)^2 (n-1) \, A\; +\; (n-2)\, x \, (2-x)\ddot{A} \, ,
\end{eqnarray}
and Equation (\ref{eq:b-purerotation}) has been used. 
Additionally, one must keep in mind the
constraint discussed above that $m=-1/2$ whenever $g \ne 0$. Equations
(\ref{eq:fundamental1}) and (\ref{eq:fundamental2}) represent 
two equations for the three variables $\rho$, $p$, and $A$. The system can be
closed by adding a relation of the kind $p=p(\rho)$, such as, e.g., a
polytropic relation $p \propto \rho^\gamma$ (discussed in
Section~\ref{sec:polytropic}). These two equations are coupled through the
$\mathcal{F}$ term, and can be combined to give an instructive form
\begin{eqnarray} \nonumber
&& \left[\frac{\dot{A}}{(n-2)A} +\frac{1-x}{x (2-x)}\right] \, \frac{U^2}{x
    \, (2-x) \,\rho} + \left[ \frac{2\, n\, p \, \dot{A}}{(n-2)A} -
    \dot{p}\right] = \\ \label{eq:fundamental} &&
  ~~~~~~~~~~~~~~~~~~~~~~~~~~~~~~~ = \rho g \left[\frac{(1-x) \dot{A}}{(n-2)A}
    -1 \right] \, ,
\end{eqnarray}
which is the fundamental equation for the situation studied in this section, since it illustrates the relations between different
parts of the system. Two different possibilities are discussed in the following
sections, namely: studying the effects of the gas pressure in the absence of
rotation and gravity ($U=0$, $g=0$; Section~\ref{sec:fff+gaspressure});
including rotation but disregarding gas pressure (zero-$\beta$ limit) and gravity
(Section~\ref{sec:lowtemperature}). A third possibility exists: disregarding gas pressure (zero-$\beta$ limit) and rotation ($U=0$), but keeping gravity. However, in this case gravity cannot be balanced along the symmetry axis or, in other words, any solution to the equations must have a singularity along the symmetry axis. So no physical solutions can be found in that case.

By analyzing the asymptotic behavior of the different quantities as $x\to0$ on the basis of the foregoing equations together with the general considerations of Section~\ref{sec:generalconsiderations} one can obtain algebraic relations limiting the admissible ranges for the exponents $m$, $n$, etc. However, given that we are going to deal with specific solutions which have more restrictive conditions on the exponents, we shall treat the restrictions individually in each subsection.

%
%


\subsection{Effect of gas pressure}\label{sec:fff+gaspressure}

The first case is a natural generalization of the force-free situation considered
in Section \ref{sec:fff}. We limit ourselves to the case without rotation
and without gravity, $U=0; g=0$. From Equation (\ref{eq:fundamental}), we
obtain 
\begin{equation}
 p = c_{p} |A|^{\frac{2 n}{n-2}}\, ,
\end{equation}
implying that the isosurfaces of $p$ and $A$ coincide. Since $A(x \to 0) \to
0$, we see that, for the pressure not to become singular toward the axis
either $n>2$ or $n \le 0$. 
Combining this with (\ref{eq:fundamental1}) one finds
\begin{eqnarray}\label{eq:generalization-LB}
&& x \, (2-x) \, \ddot{A} \,+\, (n-2) \, (n-1) \, A +\\ \nonumber
&&~~~~~~~~ +  \frac{(n-1)}{(n-2)} \; c_\phi^2 \; \sa \;|A|^{\frac{n}{n-2}} +\\ \nonumber
&&~~~~~~~~ + \frac{2n}{(n-2)} \; c_{p} \; x \; (2-x) \; \sa \; |A|^{\frac{n+2}{n-2}}\quad =\quad 0 \, ,
\end{eqnarray}
which generalizes \citet{lynden-bell1994}'s Equation
(\ref{eq:ffe}) by adding the gas pressure term (last term in the
  equation). We first study the solutions for the case 
  $n\ge 2$ and compare them with those found in section~\ref{sec:fff}. In
  that section we had had to exclude the negative-$n$ cases since
  Equation~(\ref{eq:ffe}) could not be fulfilled near the axis. 
At the end of this subsection we explore whether the extra term 
in Equation~(\ref{eq:generalization-LB}) can modify the
  negative result of the earlier section.

\begin{figure}
\includegraphics[width=0.45\textwidth]{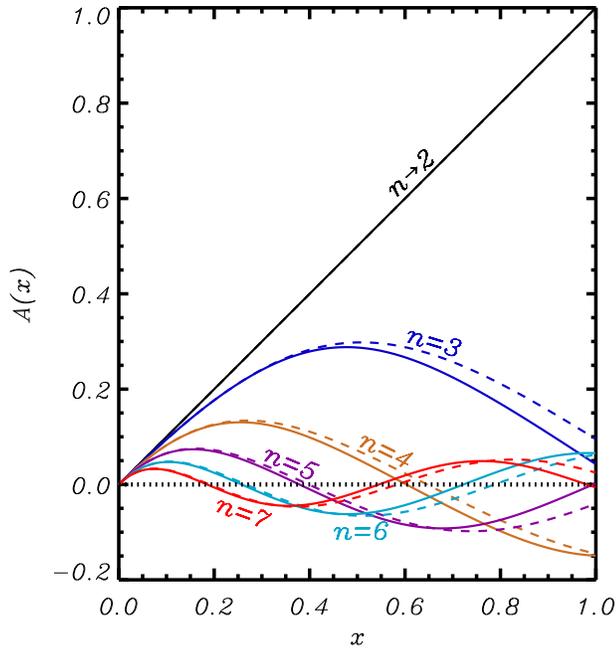}
\caption{Solid curves: solutions $A(x)$ for the case with gas
    pressure (Sec. \ref{sec:fff+gaspressure}) for different values of $n\ge2$
    with $c_\phi=5$ and $c_{p}=50$. The dashed lines are the corresponding
    force-free solutions (Sec. \ref{sec:fff}). }
\label{fig:Apress} 
\end{figure}

\begin{figure}
\begin{center}
\includegraphics[width=0.47\textwidth]{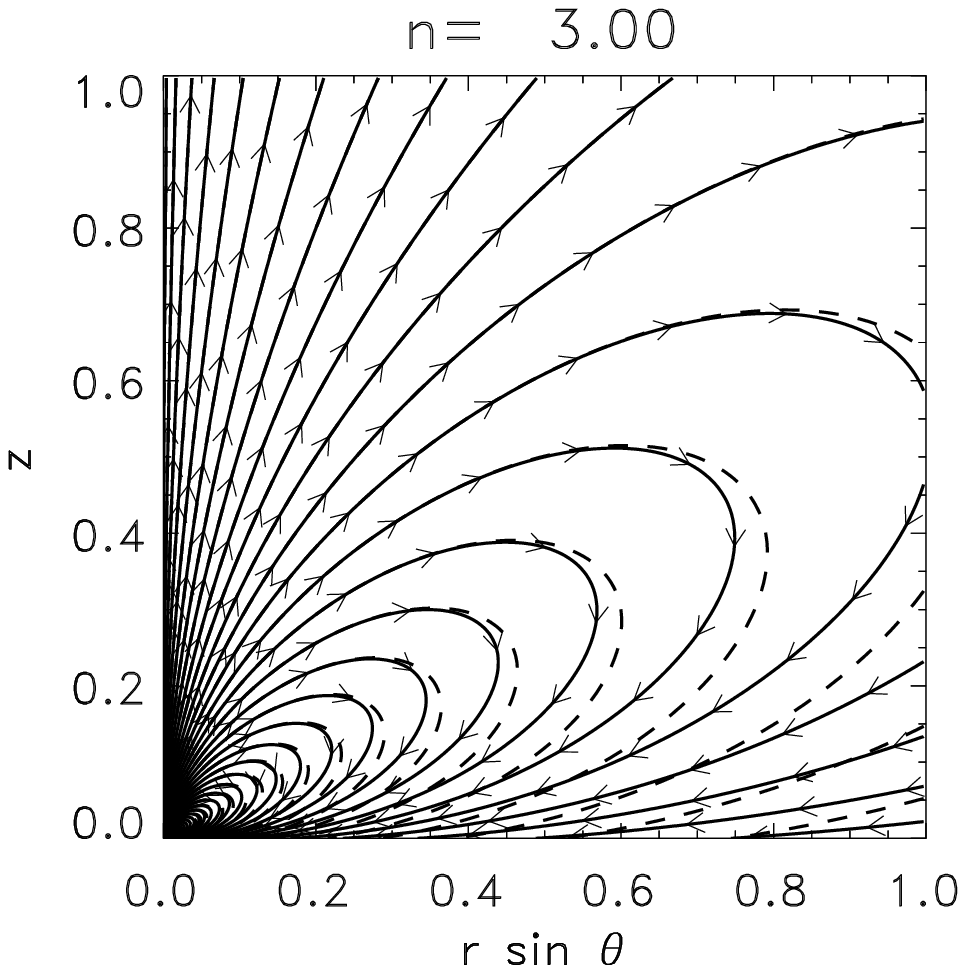}\\
\includegraphics[width=0.47\textwidth]{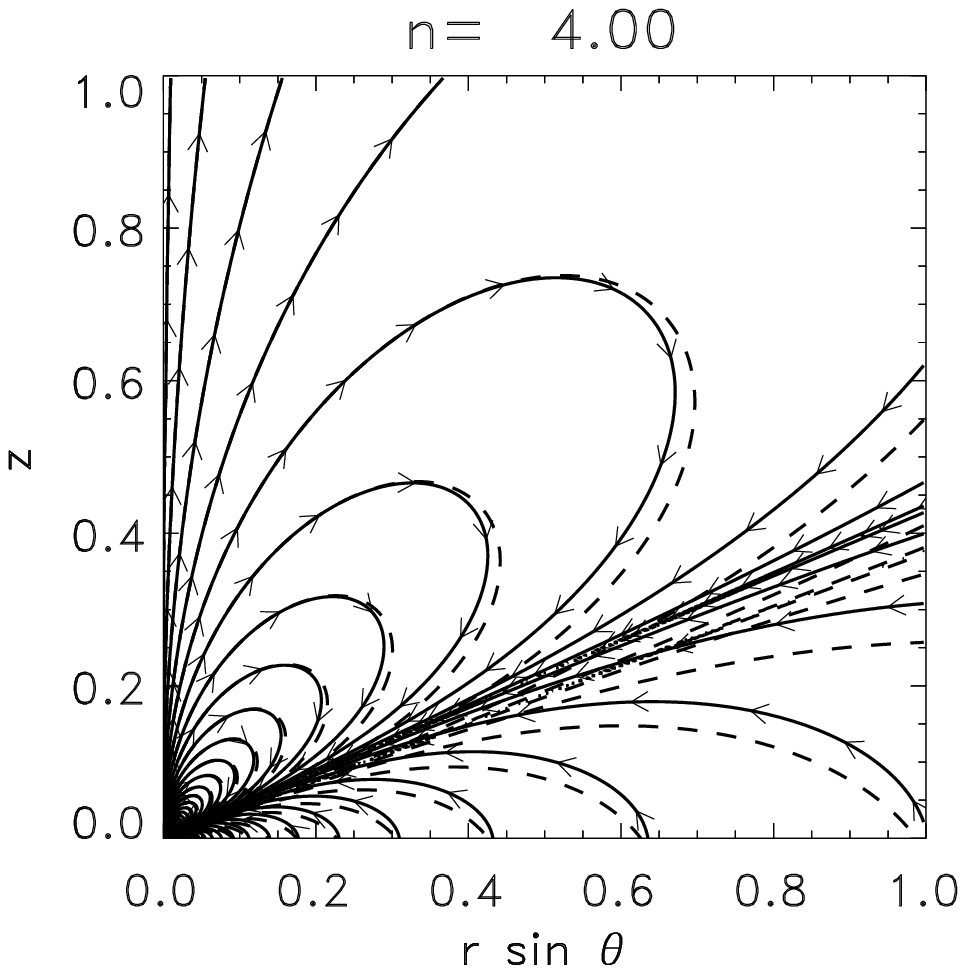}
\end{center}
\caption{Poloidal magnetic field for $n=3$, and $4$ for the situation with
  gas pressure with $c_{\phi}=5$ and $c_{p}=50$. In order to understand the
  role of the gas pressure in the equilibrium we have also plotted the
  force-free field as dashed lines.}\label{fig:poloidalfield-fff+press}
\end{figure}

For $n > 2$ we numerically integrate this equation to obtain $A(x)$ with the
boundary condition $A(0)=0$; also, for specificity, we
  arbitrarily choose $\dot{A}(0)=1$ (but could use other values if we wanted
  to explore the parameter space). In Figure \ref{fig:Apress} several solutions
  are plotted for $A$ for $n\ge 2$ alongside the force-free
  solutions (dashed) already studied ; in
  Figure~\ref{fig:poloidalfield-fff+press} the actual field lines are shown
  in the poloidal plane: in both figures the force-free solution is drawn
  solid and the non-force-free one dashed.  We see that, at least for $n=3$
  and $4$, the field lines close in on themselves nearer the central axis
  than in the force-free case. Checking with
  Figure~\ref{fig:fffA}, we see that, mathematically, this is due to the fact
  that the point $\dot{A}=0$ is reached for smaller $x$ in the potential than in
  the force-free case: since the field lines are isolines of $\hat{A} =
  A(x)/r^{n-2}$ (as follows from Equation~\ref{eq:full_magnetic_potential}),
  along each field line the points where $\dot{r}=0$ and $\dot{A}= 0$ must
  coincide: the turning point of the field line (where $\dot{r}=0$)
  occurs at smaller $x$ for the non force-free case. The physical reason for
  this behaviour, on the other hand, can be seen as follows: the pressure
  gradient must be zero along each individual field line, since there is no
  Lorentz force component in that direction to compensate for it. So, looking
  at the plane drawn in Figure~\ref{fig:poloidalfield-fff+press}, the
  pressure gradient is parallel to the gradient of $\hat{A}$. The value of
  $\hat{A}$, in fact, decreases when jumping outward from one field line to
  the next, and so must the pressure. Hence the pressure gradient force
  points outward, and the Lorentz force must be reinforced to compensate for
  it: the extra curvature of the poloidal field lines helps in doing that; by
  studying the Lorentz force term ${\cal F}(A,\ddot{A})$ in
  Equations~(\ref{eq:fundamental1})--(\ref{eq:F_definition}), one can
  conclude that both the azimuthal and poloidal field components help to
  reinforce the Lorentz force in this case.
From the Figure  we also see that, for these $n\ge 2$ solutions, the influence of the
  pressure is most marked at large $x$, i.e., large $\theta$, whereas it is
  unimportant near the axis.

We finish by considering the $n < 0$ solutions. The reason why there were no such solutions in the force-free case
  of Section~\ref{sec:fff} was that the magnetic pressure term $x(2-x)
\ddot{A}$ could not be balanced by any other term in
Equation~\ref{eq:ffe}. Now we see that the new extra term, the gas pressure
term, is in no position to compensate for it, either, so the negative-$n$
solutions must be excluded also here. The solutions in this section therefore have closed field lines and no open field lines are allowed. Thus, no
collimated structures appear with the new gas pressure term.

\subsection{Effect of rotation}\label{sec:lowtemperature}

To explore the effects of rotation, we here go to the
  zero-$\beta$ (i.e., no gas-pressure) limit without gravity; from
  (\ref{eq:fundamental}) we then see that

\begin{equation}\label{eq:purerotation-lowt-a0equation}
\frac{\dot{A}}{A} = (2-n) \frac{1-x}{x(2-x)} \, ,
\end{equation}
which is easily integrable to
\begin{equation}\label{eq:purerotation-sol-A}
A(x)=c_A \left[ x (2-x) \right]^\frac{2-n}{2} \, ,
\end{equation}
with $c_A$ an integration constant that coincides
  with $A(1)$. Given the definition of $x$, $A(x)$ is therefore monotonic and
  simply proportional to $\sin^{2-n}\theta$. We are free to select 
its global sign, and in the following assume $A(x)>0$. From
(\ref{eq:purerotation-sol-A}), for $A$ to 
vanish at the axis,  $(2-n)/2\ge~1$, i.e., $n\le~0$. As we
shall see, $n=0$ does not satisfy the asymptotic conditions for $b$ at the
axis and so the allowed values are $n<0$ : it is
interesting to see that in this case solutions with a collimated structure
are found as in the case of a potential field (Sect. \ref{sec:potential-field}).

\begin{figure}
\includegraphics[width=0.45\textwidth]{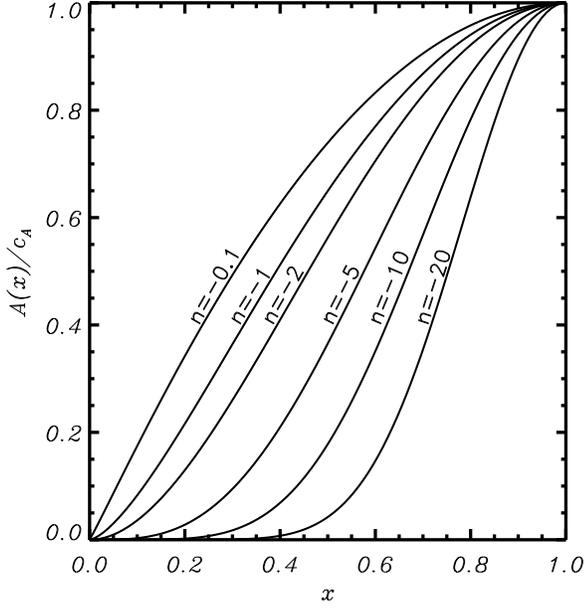}
\caption{Pure rotation with no gravity and low temperature, showing the lowest-order magnetic potential $A(x)/c_A$ for several values of $n<0$. \label{fig:purerotation-A-solutions}}
\end{figure}

In Figure \ref{fig:purerotation-A-solutions} several normalized solutions
$A(x)/c_A$ are plotted for different values of $n$. For all the allowed $n$ values the derivative of $A$ at the origin is zero. For
$n=-0.1$ the function rapidly grows very close to the origin. For smaller
values of $n$ the growth is less pronounced, and for $n=-20$ the function is
close to zero for a large portion of the domain.

\begin{figure}
\centering\includegraphics[width=0.47\textwidth]{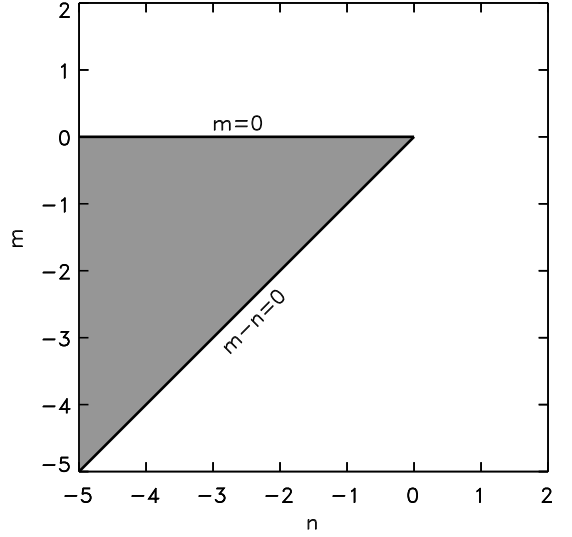}
\caption{Allowed values of $m$ and $n$ for pure rotation and a zero-$\beta$ plasma given by the axisymmetric conditions. In the white area at least one of the conditions is not fulfilled. The shaded area gives the allowed values for solutions of Equation (\ref{eq:purerotation-sol-A}). The solid thick lines are plotted to show that these values are included in the region of validity.\label{fig:allowed_mn_purerotion}}
\end{figure}

From Equations (\ref{eq:u-purerotation_2}) to (\ref{eq:F_definition}) we find\
\begin{equation}\label{eq:purerotation-simple-rho} 
\rho(x)= c_\rho \, \left[ x \, (2-x) \right]^{m-n} \, ,
\end{equation}
where
\begin{equation}\label{eq:purerotation-rho-coef}
c_\rho \equiv \frac{(1-n) c_A^{\frac{2}{n-2}}- n \, (n-2)^2 \, c_\phi^{-2}  }{c_U^2\, c_\phi^{-2} \, c_A^\frac{2(3+m-n)}{n-2}}  \, .
\end{equation}
Combining Equations (\ref{eq:u-purerotation_2}), (\ref{eq:b-purerotation}), (\ref{eq:purerotation-sol-A}), and (\ref{eq:purerotation-simple-rho}) we obtain
\begin{eqnarray}\label{eq:u-purerotation_2-simple} 
U(x) &=& c_U \, c_\rho \, c_A^{\frac{1+m}{n-2}} \left[ x(2-x) \right]^{\frac{1}{2}(m-2n+1)} \, ,\\ \label{eq:b-purerotation-simple} 
b(x) &=&
  c_\phi \, c_A^\frac{n-1}{n-2} [x(2-x)]^\frac{(1-n)}{2}  \, .
\end{eqnarray}
The exponents of the square bracket terms in
  (\ref{eq:purerotation-simple-rho}),
  (\ref{eq:u-purerotation_2-simple}) and (\ref{eq:b-purerotation-simple})
  directly give the asymptotic behavior 
  of $\rho(x)$, $U(x)$ and $b(x)$ when $x\to~0$. In
  Section~\ref{sec:generalconsiderations} we had given the symbols $\vvv$,
  $\nu$ and 
  $\omega$, respectively, for those exponents and seen that they must fulfill
  relations (\ref{eq:omega}) -- (\ref{eq:nu_sigma}), additionally to the
  condition $\vvv \ge 0$ for $\rho$ not to be singular at the
  axis. Putting all that together we conclude that $n \le m <0$.
 In Figure \ref{fig:allowed_mn_purerotion} the region of allowed $m$ and
$n$ is plotted as a shaded area. For $m$ and $n$ on the line $m-n=0$ (thick
 line), the density $\rho(x)$ is independent of $x$ according to
 Equation (\ref{eq:purerotation-simple-rho}).

From Equation (\ref{eq:purerotation-rho-coef}) we see that the four
coefficients $c_\rho, c_A, c_U$, and $c_\phi$ are related. This implies that
we have freedom to fix three of the four coefficients and the remaining one
is set by this relation. For example, fixing the magnetic structure (with
$c_A$ and $c_\phi$) and a given rotation velocity ($c_U$), the density
($c_\rho$) accommodates accordingly. Similarly, for a given magnetic field
structure and density the plasma will rotate with a given profile. For the
allowed values of $n<0$ the coefficient $c_\rho$ is always positive as
expected.

The structure of the magnetic field in this situation is quite
simple. Calling $\rperp$ the cylindrical radius, i.e., $\rperp =
r\,\sin\theta = r \sqrt{x(2-x)}$, and $\Bperp$ the field component in that
direction, $\Bperp = B_{r}\, \sin \theta  + B_{\theta}\,\cos \theta $,
we see that for all allowed values of $n$, the poloidal field 
is everywhere aligned with the $z$-axis: 

\begin{eqnarray}
\Bperp &=& 0\,, \label{eq:nocylfield}\\
B_{z} &=& B_{r}\,\cos \theta \,  - \,B_{\theta} \sin \theta \,\;=\;\, (2-n)\,
c_A\,\rperp^{-n} \;. \label{eq:purerotation-bz-x} 
\end{eqnarray}
To obtain that result one must use Equations
(\ref{eq:generalconsideration-br}), (\ref{eq:generalconsideration-btheta}) 
and (\ref{eq:purerotation-sol-A}). When written out in their full dependence
with $r$ and $\theta$, the other magnitudes 
$B_\phi$,  $\rho$, and $v_\phi$ can also be seen to depend $\rperp$ only,  as
follows:
\begin{eqnarray}\label{eq:purerotation-bphi-x}
B_{\phi}(r,\theta)&=& c_\phi \, c_A^\frac{n-1}{n-2} \rperp^{-n}\, , 
              \\ 
\noalign{\vspace{2mm}}
\label{eq:purerotation-rho-x}
\rho (r,\theta)&=& c_\rho \, \rperp^{2(m-n)}\, ,
              \\ 
\noalign{\vspace{2mm}}
\label{eq:purerotation-vphi-x} 
v_{\phi} (r,\theta)&=& c_U \, c_A^\frac{1+m}{n-2} \, \rperp^{-m} \, . 
\end{eqnarray}
The system therefore has cylindrical symmetry; the index
  $n$ can take any negative value, indicating that the magnetic field always
  increases with the distance from the rotation axis. The
  index $m$ is constrained by the condition $0 > m\ge n$ (see
  Fig. \ref{fig:allowed_mn_purerotion}). Thus, the rotation velocity $v_\phi$
  can increase ($m<0$) with distance from the rotation axis but never faster
  than the magnetic field. The density [Equation~(\ref{eq:purerotation-rho-x})] depends on the $m-n$ index, which is always positive (Fig. \ref{fig:allowed_mn_purerotion}). For $m-n>0$ the density vanishes at the rotation axis and increases with distance from this axis. The profile of $\rho$ complements the constraint on $v_\phi$ in that $\rho v_\phi^2$ has the same dependence on $\rperp$ as $B_\phi^2$, expected to maintain dynamical equilibrium.

\begin{figure}
\includegraphics[width=0.45\textwidth]{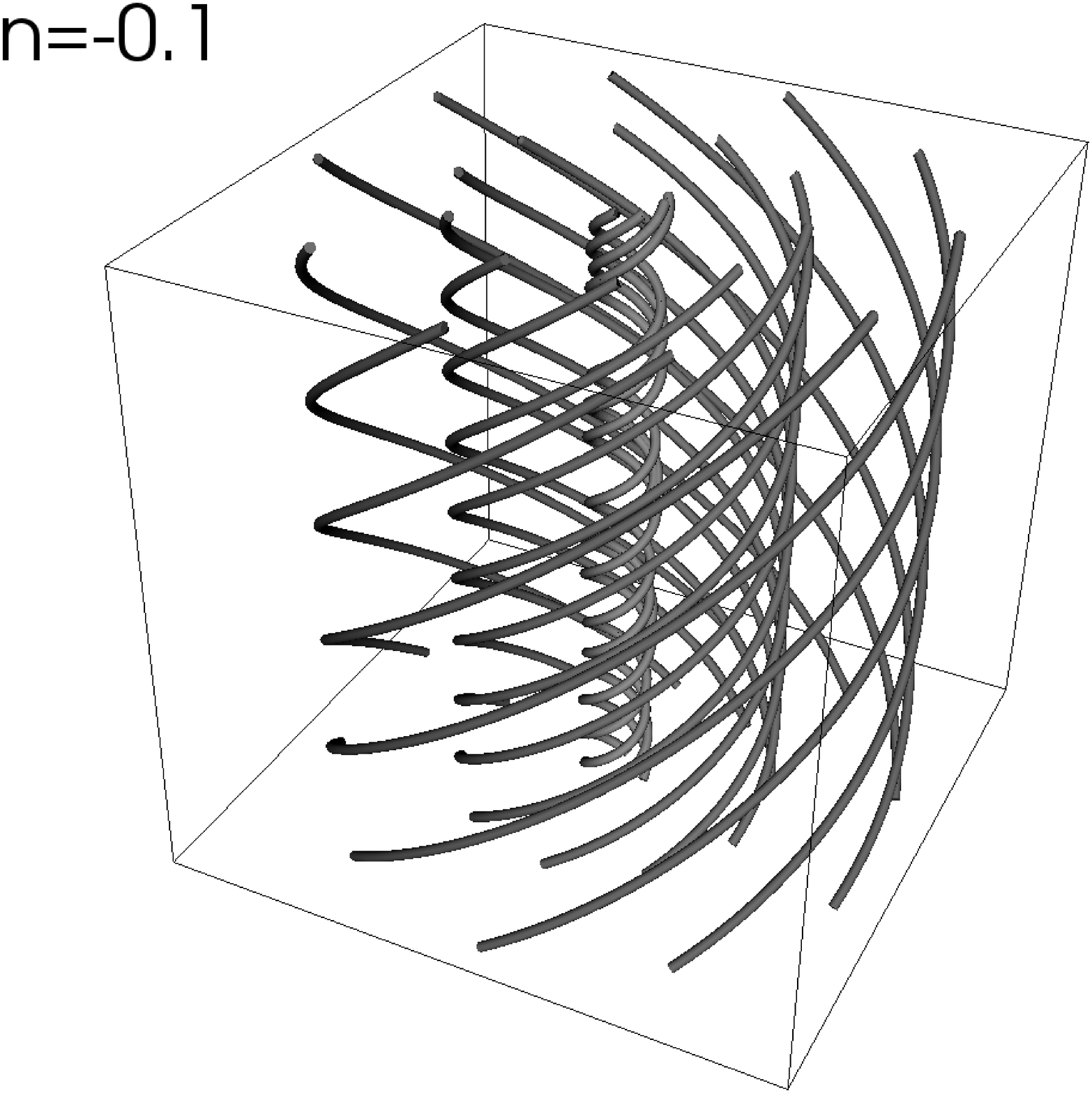}
\includegraphics[width=0.45\textwidth]{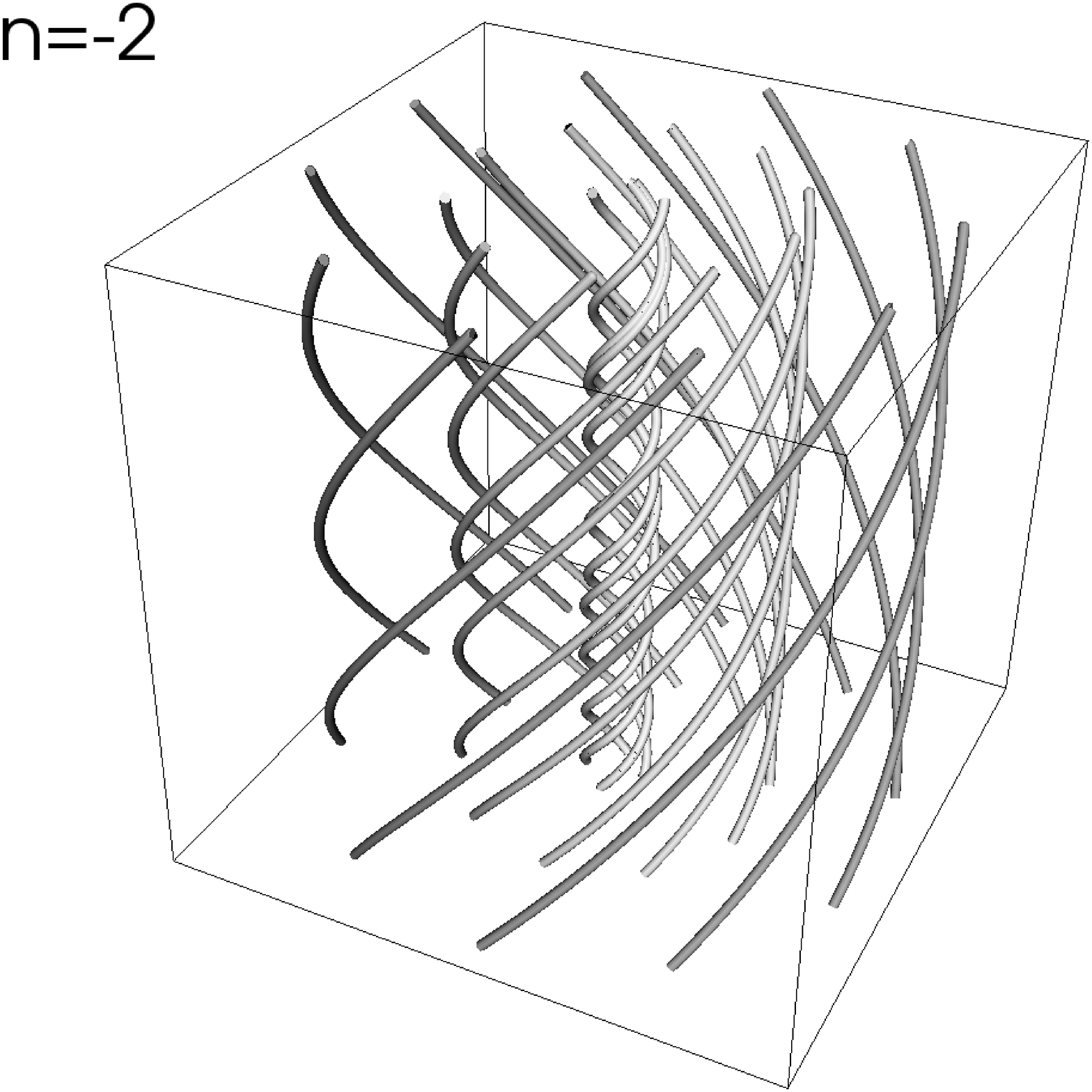}
\caption{Pure rotation in the zero-$\beta$ limit and with no gravity, showing the field lines for the magnetic field of Equations (\ref{eq:purerotation-sol-A}) and (\ref{eq:b-purerotation-simple}). The color code indicates the strength of the magnetic field from white to black. The magnetic field for $n=-0.1$ and $n=-2$ solutions are shown in the top and bottom panel, respectively. In both cases $c_\phi=2$ and $c_A= 2$.\label{fig:3d-purerotation-solution}}
\end{figure}

Figure \ref{fig:3d-purerotation-solution} shows the three-dimensional
magnetic field lines. The cylindrical symmetry of the system
  is apparent.  
In the cases shown the magnetic field vanishes at the rotation
axis: for $n=-0.1$ (top panel) the magnetic field is more uniform
than in the case of $n=-2$ (bottom panel), as follows from the simple
power-laws of (\ref{eq:purerotation-bz-x}) and
(\ref{eq:purerotation-bphi-x}).

In general, the field line inclination is given by the
ratio of $B_{\phi}/B_{z}$, which, after using Equations
(\ref{eq:purerotation-bz-x}) and (\ref{eq:purerotation-bphi-x}), becomes 
\begin{equation}\label{eq:purerotation-twist}
\frac{B_{\phi}}{B_{z}}=\frac{1}{2-n} \, c_\phi \, c_A^{\frac{1}{n-2}} \, ,
\end{equation}
and so is independent of position. 
The uniformity of the ratio $B_{\phi}/B_z$ matches the visual impression in 
Figure \ref{fig:3d-purerotation-solution}. The maximum inclination is 
reached for $n=0$ and decreases when $n$ decreases for constant $c_\phi$ and $c_A$.

The physical quantities in the foregoing solutions ($B, \rho, v_\phi$) all
increase with the cylindrical radius $\rperp$. Hence their application must
necessarily be limited to a finite region of space, i.e., beyond a given
radius they should be matched with an external solution via adequate boundary
conditions. Wherever such matching can be carried out successfully, these
solutions could provide useful insights into the nature of rotating MHD
structures.

\vspace{1cm}
\section{Discussion and Conclusions}
In this work we have set up models for MHD structures that are axisymmetric
and steady state. We have sought self-similar solutions of the form
$f(\theta)/r^n$ with each variable having a different function $f(\theta)$
and power, $n$. We have proceeded from the simplest situation to ones of
increasing complexity.

The first system we consider is a potential field , so necessarily with no
electric current and no magnetic twist. We have found analytical solutions
for different values of $n$. For $n>0$ the origin of coordinates is a source
of magnetic field and its strength declines with distance. In contrast, for
$n<0$ the origin is a null point and the magnetic field strength
increases. For $n=0$ the magnetic field is uniform in space and parallel to
the $z$-axis. For $0 < n < 2$ the field lines curve towards the the $z$-axis
and for $n=2$ the field lines are straight, but radiate in all directions
from the origin. For $n > 2$ they curve away from the $z$-axis and form
loops, closing back down to meet the horizontal axis. One of the most
sets of interesting solutions is the set with $n<0$. Here the magnetic field at
the origin vanishes and so there is a null point there with no closed
lines. For $n=-0.5$ the field lines are curved with respect to the vertical
field forming a collimated magnetic structure. For $n=-1$ the magnetic field
is identical to a first-order null point with a typical X shape at the
origin.  For $n=-2$, the solution has a fan in the domain and for $n<-2$ the
complexity increases with an increasing number of fans. This simple solution
reveals that it is necessary to have $n$ negative to form a collimated
structure around the axis. The collimated structure consists of open field
lines that accumulate around the axis.

The second system we consider is force-free and includes twist. There exist
solutions here only for $n\ge 2$ where the field forms loops starting at the
centre returning again to the bottom surface as in the potential
situation. However, the presence of the azimuthal component twists the loops
around the axis. The case with $n\ge2$ is similar to the case studied by
\citet{lynden-bell1994} but with different boundary conditions. In this case,
the $n<0$ solutions are forbidden and no collimated magnetic structures
appear: the extra force associated with the new azimuthal term
cannot be balanced by the poloidal magnetic pressure and tension at the axis.

We then increased the complexity of the system by considering a more general
situation, namely non-force free cases but still with no poloidal flows. This
allows us to study structures with non-vanishing gas pressure,
rotation, or gravity. The first case is a system with magnetic field and gas
pressure but without rotation and gravity. This extends Lynden-Bell\&Boily's force-free solutions by adding gas pressure. The
solutions have field lines that are more highly curved than in the force-free
solutions: the extra pressure gradient must be balanced by the magnetic
force. Also, the gas pressure is more important close to the horizontal
axis. The allowed solutions have $n \ge 2$ indicating that the extra pressure
cannot balance the magnetic pressure at the axis when $n<2$. As in the
non-potential force-free situation the resulting structure is not collimated
around the axis.

For rotating  zero-$\beta$ plasma structures we find analytical solutions for
the magnetic field, rotational velocity, and density fields. They all possess a
cylindrical geometry and depend solely on distance from the rotation axis but
they exist only for $n<0$. The magnetic field has twist that depends on the
index $n$. The density and rotation velocity increase with the distance from
the centre. The new velocity term introduces a centrifugal force, that
contributes to the balance of the Lorentz force associated with the poloidal
and azimuthal components of the magnetic field. In fact, the poloidal field
consists of straight field lines indicating that the poloidal tension
vanishes, and the centrifugal force balances the inward magnetic pressure and
azimuthal tension forces.

Summarizing, in this paper we have formulated general equations for the
steady-state ideal MHD problem assuming axisymmetry and self-similarity and
including flows, gas pressure and gravity. Also, a number of solutions have
been calculated including potential, force-free and non-force free ones.
The results can be used as a starting point for future developements, such as, for example, including poloidal flows in the solutions.  In \citet{luna2015} we showed that the combination of magnetic twist and poloidal flow can induce
a force along the axis of the structure. This is an interesting scenario for
producing jets and supporting the cool plasma against
gravity in the solar corona.

\section*{Acknowledgements}
Support by the Spanish Ministry of Economy and Competitiveness through
project AYA2014-55078-P is acknowledged. M.~L.~also acknowledges support from
the International Space Science Institute (ISSI) to the Team 374 on ``Solving
the Prominence Paradox'' led by Nicolas Labrosse. ERP is most grateful for warmth and hospitality during his visits to the IAC.

\end{document}